\documentstyle[amssymb,psfig]{mn2e}

\title[CIRPASS spectroscopy of massive star clusters in NGC
1140]{CIRPASS near-infrared integral-field spectroscopy of massive star
clusters in the starburst galaxy NGC 1140}

\author[R. de Grijs et al.]{R. de Grijs,$^{1,2}$\thanks{E-mail:
R.deGrijs@sheffield.ac.uk} L. J. Smith,$^3$ A. Bunker,$^{1,4}$
R. G. Sharp,$^{1,5}$ J. S. Gallagher {\sc iii},$^6$
\newauthor P. Anders,$^7$ A. Lan\c{c}on,$^8$ R. W. O'Connell$^9$ and I. R.
Parry$^1$
\\ 
$^1$ Institute of Astronomy, University of Cambridge, Madingley Road,
Cambridge CB3 0HA\\
$^2$ Department of Physics \& Astronomy, The University of Sheffield, Hicks
Building, Hounsfield Road, Sheffield S3 7RH \\
$^3$ Department of Physics \& Astronomy, University College London,
Gower Street, London, WC1E 6BT\\
$^4$ Present address: School of Physics, University of Exeter, Stocker
Road, Exeter, EX4 4Q\\
$^5$ Anglo-Australian Observatory, P.O. Box 296, Epping, NSW 1710, 
Australia\\
$^6$ Astronomy Department, University of Wisconsin-Madison, 475 N.
Charter St., Madison, WI 53706, USA \\
$^7$ Universit\"atssternwarte, University of G\"ottingen,
Geismarlandstr. 11, 37083 G\"ottingen, Germany \\
$^8$ Observatoire Astronomique, Universit\'e L. Pasteur \& CNRS:
UMR 7550, 11 rue de l'Universit\'e, 67000 Strasbourg, France \\
$^9$ Astronomy Department, University of Virginia, P.O. Box 3818,
Charlottesville, VA 22903, USA
}

\date{Received date; accepted date}
\pubyear{2003}

\begin{document}
\maketitle

\begin{abstract}
We analyse near-infrared integral field spectroscopy of the central
starburst region of NGC 1140, obtained at the Gemini-South telescope
equipped with CIRPASS. Our $\sim 1.45 - 1.67 \mu$m wavelength coverage
includes the bright [Fe {\sc ii}]$\lambda 1.64 \mu$m emission line, as
well as high-order Brackett (hydrogen) lines. While strong [Fe {\sc
ii}] emission, thought to originate in the thermal shocks associated
with supernova remnants, is found throughout the galaxy, both Br 12--4
and Br 14--4 emission, and weak CO(6,3) absorption, is predominantly
associated with the northern starburst region. The Brackett lines
originate from recombination processes occurring on smaller scales in
(young) H{\sc ii} regions. The time-scale associated with strong [Fe
{\sc ii}] emission implies that most of the recent star-formation
activity in NGC 1140 was induced in the past $\sim 35-55$ Myr. Based
on the spatial distributions of the [Fe {\sc ii}] versus Brackett line
emission, we conclude that a galaxy-wide starburst was induced several
tens of Myr ago, with more recent starburst activity concentrated
around the northern starburst region.\\
This scenario is (provisionally) confirmed by our analysis of the
spectral energy distributions of the compact, young massive star
clusters (YMCs) detected in new and archival broad-band {\sl Hubble
Space Telescope} images. The YMC ages in NGC 1140 are all $\lesssim
20$ Myr, consistent with independently determined estimates of the
galaxy's starburst age, while there appears to be an age difference
between the northern and southern YMC complexes in the sense expected
from our CIRPASS analysis. Our photometric mass estimates of the NGC
1140 YMCs, likely upper limits, are comparable to those of the
highest-mass Galactic globular clusters and to spectroscopically
confirmed masses of (compact) YMCs in other starburst galaxies. Our
detection of similarly massive YMCs in NGC 1140 supports the scenario
that such objects form preferentially in the extreme environments of
interacting and starburst galaxies.
\end{abstract}


\section{Introduction}
\label{intro.sec}

The southern starburst dwarf galaxy NGC 1140 (Mrk 1063) is remarkable
in the number of compact, young, massive star clusters (YMCs) that it
has formed recently in the small volume of its bright emission-line
core (roughly $600 \times 200$ pc at the galaxy's
distance\footnote{based on the galaxy's heliocentric velocity,
corrected for the Virgocentric flow, and assuming H$_0 = 70$ km
s$^{-1}$ Mpc$^{-1}$; adopted from the {\sl HyperLeda} database at {\tt
http://www.univ-lyon1.fr/}} of $(m-M)_{\rm kin} = 31.49$ mag or $D =
20$ Mpc). Based on H{\sc i} and optical observations, and comparison
with models (Hunter, van Woerden \& Gallagher 1994b, hereafter H94b),
it exhibits the characteristics of a recent merger event, which has
presumably induced the YMC formation in its centre; the galaxy is
unusually bright ($M_V^0 = -20.76$; de Vaucouleurs et al. 1991,
hereafter RC3) for its dwarf (or amorphous) morphological type, and
shows low-level shells and filaments in its H{\sc i} distribution
(H94b), which are presumably remnants of a past interaction.  Within
3.3 Holmberg radii $(R_{\rm H}$, the radius where the {\it V}-band
surface brightness reaches $\mu_V = 26.4$ mag arcsec$^{-2}$; at $\mu_V
= 25$ mag arcsec$^{-2}$ the galaxy measures about $9.9 \times 5.2$
kpc), it contains a total mass of $1 \times 10^{11} {\rm M}_\odot$,
based on its H{\sc i} rotation curve; $7.5 \times 10^9 {\rm M}_\odot$
of this mass is in the form of H{\sc i}. The total mass contained
within $1 R_{\rm H}$ is $1.4 \times 10^{10} {\rm M}_\odot$ (H94b).

The formation of compact YMCs seems to be a key feature of intense
episodes of recent, active star formation. They have been identified,
mostly with the {\sl Hubble Space Telescope (HST)}, in several dozen
galaxies, often interacting or undergoing active starbursts (see,
e.g., Holtzman et al. 1992, O'Connell, Gallagher \& Hunter 1994, de
Grijs, O'Connell \& Gallagher 2001, Whitmore 2003, and references
therein). Their sizes, luminosities, and masses -- usually based on
broad-band optical observations -- are generally consistent with
theoretical expectations for young globular clusters (see also de
Grijs, Bastian \& Lamers 2003b). However, they cover a wide range of
masses and luminosities, and also include those of the most extreme
supermassive and superluminous clusters. YMCs are also important as
probes of their host galaxy's star formation history, its chemical
evolution, the stellar initial mass function (IMF), and other physical
characteristics in starbursts.  This is so because each cluster
approximates a coeval, single-metallicity, simple stellar
population. Such systems are among the simplest to model, and their
ages and metallicities and, in some cases, IMFs can be estimated from
their integrated spectra (Ho \& Filippenko 1996a,b, Gallagher \& Smith
1999, Smith \& Gallagher 2001, Mengel et al. 2002).

\subsection{NGC 1140: A disturbed appearance across the board}

Observations in optical broad-band, H$\alpha$ narrow-band, and the
H{\sc i} radio continuum are consistent with the interpretation that
NGC 1140 is in the final stages of a galactic merger (e.g., Hunter et
al. 1994a, hereafter H94a; H94b). At faint optical light levels it
shows multiple, misaligned shell-like structures. These are
reminiscent of the shells associated with some elliptical galaxies
that are thought to be remnants of a past tidal encounter (e.g.,
Schweizer \& Seitzer 1988, Hernquist \& Spergel 1992). The galaxy is
also unusual in its content and overall morphology: its optical
appearance is dominated by a supergiant H{\sc ii} region encompassing
most of its centre, with an H$\alpha$ luminosity equivalent to that
produced by $\sim 10^4$ OB stars (H94a), far exceeding that of the
giant H{\sc ii} region 30 Doradus in the Large Magellanic Cloud (LMC;
H94b). Its gas fraction, gas-to-luminosity ratio, and H{\sc i}
velocity dispersion are unusually high, and its H$\alpha$ velocity
profiles remarkably broad, by roughly an order of magnitude, for the
mid-spiral type galaxy one would expect, based on its total mass
(H94b).  Although NGC 1140 has a small dwarf companion galaxy (which
might suggest that gravitational interactions with this companion may
have induced part of the recent active star cluster formation), these
observations led H94b to conclude that the galaxy is most likely the
product of the accretion of a low surface brightness gas-rich
companion by a relatively normal mid-type spiral galaxy of roughly
equal mass in the last $\sim 1$ Gyr. This picture is consistent with
the low metallicities (see Section \ref{hst1.sec}) and complex H{\sc
i} morphology (including a warped disc) and velocity field.

\subsection{The massive star cluster population}

The presence and characteristics of the central supergiant H{\sc ii}
region in NGC 1140 are consistent with simple conceptual models of
enhanced central star formation associated with a tidal encounter (see
H94b). Its core contains six blue, very luminous compact
YMCs\footnote{H94a show convincingly that these objects are much more
likely to be YMCs than either luminous red supergiants or objects
similar to the most luminous unstable stars} -- with ages $\lesssim
15$ Myr (H94a) -- such as commonly found in regions of intense star
formation.  These are comparable to or brighter than Galactic globular
clusters (GCs) at similar ages, and thought to be entirely responsible
for the ionisation of the supergiant H{\sc ii} region in NGC 1140
(H94a). It appears that most of the recent star formation in NGC 1140
has occurred in, or is associated with, the YMCs: while its overall
average recent star formation rate, $\dot{M} \sim (1-6) {\rm M}_\odot$
yr$^{-1}$ (based on a variety of tracers of the recent star formation
rate and depending on -- among others -- the slope and high-mass
cut-off assumed for the IMF; e.g., Storchi-Bergmann, Calzetti \&
Kinney 1994, H94b, Calzetti 1997, Takagi, Arimoto \& Hanami 2003), is
similar to the equivalent rates of the significantly more massive
giant irregular and spiral galaxies, the average recent star formation
rate {\em in the YMCs alone} is implied to be near 1 M$_\odot$
yr$^{-1}$ (H94a), if their IMFs are similar to the solar neighbourhood
IMF.

In Section \ref{observations.sec} we describe in detail both the
broad-band {\sl HST} imaging observations of the few dozen compact
YMCs (i.e., the six very bright clusters discussed above, in addition
to a larger population of less luminous compact star clusters) in NGC
1140's central starburst, and our near-infrared (NIR) integral-field
spectroscopy. We then apply sophisticated age dating techniques to the
{\sl HST} observations in Section \ref{hst.sec} and discuss the
spectroscopic evidence for a possible age gradient (see Section
\ref{cirpass.sec}). These results are discussed in the context of NGC
1140's recent star formation history and its current supernova rate
(SNR) in Sections \ref{context.sec} and \ref{snr.sec}, respectively,
upon which we summarise our main results and conclusions in Section
\ref{summary.sec}.

\section{Observations and data reduction}
\label{observations.sec}

\subsection{Spectroscopy}
\label{spectroscopy.sec}

The {\it H}-band spectroscopy of NGC 1140 was performed with the Cambridge
Infrared Panoramic Survey Spectrograph (CIRPASS; Parry et al. 2000) at the
f/16 focus on the 8-m Gemini-South telescope, at Cerro Pachon in Chile.
CIRPASS is a NIR fibre-fed spectrograph, connected to a 490-element integral
field unit (IFU). The variable lenslet scale was set to 0.36 arcsec diameter,
and the hexagonal lenslets are arranged in the IFU to survey an approximately
rectangular area of $13.0 \times 4.7$ arcsec, which provides sufficient
sampling of the unusually bright diffuse light in NGC 1140. The detector is a
1k$\times$1k Hawaii-I HgCdTe Rockwell array. CIRPASS can operate in the range
$0.9 - 1.8 \mu$m (but with best results in the range $1.0 - 1.67 \mu$m), and a
400l/mm grating was used which produced a dispersion of 2.25{\AA} pixel$^{-1}$
and a coverage of $0.22 \mu$m for each grating setting.

The grating was tilted to place the wavelength range $1.45 - 1.67 \mu$m on the
detector, covering most of the {\it H}-band transmission window. A filter at
$1.67 \mu$m blocked out redder wavelengths to reduce the dominant NIR
background contribution. The detector pixels do not quite critically sample
the spectral resolution (unresolved sky lines have FWHM = 1.7 pixels, or
3.8{\AA}), and the resolving power is $\lambda / \Delta\lambda_{\rm FWHM} =
3500$.

The observations were made on the night of UT 6 August 2002 during the
Director's discretionary time instrument commissioning/IFU-demonstration
science on Gemini-South. The observations were taken in non-photometric
conditions, with a seeing of $\approx 1.0$ arcsec FWHM, and spanned the
airmass range $1.1-1.4$.

We obtained three 40-minute exposures of the galaxy, and one 40-minute
offset sky exposure (offset by 200 arcsec). The long axis of the IFU
was aligned along a north--south axis. In between each of the
individual 40-minute exposures, we dithered by 2 lenses, or $\sim
0.72$ arcsec. Although we had noticed that our last 40-minute exposure
was affected by high winds (and therefore blurred), during the
analysis stage we found that the guiding of one of the other exposures
had experienced problems as well due to wind shake.  Thus, the CIRPASS
results discussed in this paper are based on the single 40-minute
exposure obtained under the best available conditions.

For each individual integration of 40 minutes per pointing, the array
was read out non-destructively every 10 minutes (i.e., 5 times per
pointing, including an initial read to reset the detector). Each of
these ``loops'' comprised 10 multiple reads of the detector, which
were averaged to reduce the read-out noise (30$e$- per read) by
$\sqrt{10}$. The average of each loop was subtracted from the average
of the next loop of non-destructive reads, to form 4 sub-integrations
of 10 minutes at the same telescope pointing. Comparison of these
sub-integrations enabled cosmic ray strikes to be flagged, and a
combined frame of 40 minutes was produced by summing the four
sub-integrations, ignoring pixels affected by cosmic rays. An initial
subtraction of sky and dark current was performed using the offset sky
frame.  Known bad pixels were also interpolated at this stage.

The 490 fibres span the 1k detector, with $2\times2$ pixels per
fibre. A lamp was used to illuminate 10 calibration fibres immediately
before the observations, in order to accurately determine the position
of the fibres on the array and to focus the spectrograph at the
desired wavelength range. There is significant cross-talk between
adjacent fibres and we use an optimal-extraction routine\footnote{The
CIRPASS data reduction software is available from {\tt
http://www.ast.cam.ac.uk/$\sim$optics/cirpass/docs.html}}
(R.A. Johnson et al., in prep.) to determine the spectrum of each
fibre, thereby for each pixel on the array solving the contribution of
flux from adjacent fibres.

Immediately after the science integrations, spectral flat fields were
obtained (by means of exposures of the illuminated dome), and these
were also optimally extracted and their average flux normalised to
unity. The extracted science data was then flat-fielded through
division by the extracted normalised dome lamp spectra, thus
calibrating the response of each individual fibre.

Wavelength calibration was achieved by matching the OH lines in the
sky exposure with the calibrated sky lines in the Maihara et
al. (1993) atlas.  Again, the individual fibre spectra were optimally
extracted, and for each fibre a cubic fit was performed to the
centroids of the dozen brightest telluric OH lines, leaving
r.m.s. residuals of 0.2\AA . The 490 fibre spectra stacked on the slit
were rectified (mapped to the same wavelength) with a fifth-order
polynomial transformation in $x$ (i.e., along the wavelength axis) and
$y$ (because the wavelength distortion changes as a function of fibre
position). This transformation was applied to the sky-subtracted,
extracted, flat-fielded science frames. A higher-order background
subtraction was then applied to the rectified data, to remove sky
residuals from the beam switching caused by variation in the OH-line
intensity. The individual rectified, background-subtracted fibre
spectra composing the two-dimensional (2D) intermediate spectrum
partially shown in Fig. \ref{2d.fig} (in Section \ref{cirpass.sec}
below), were subsequently rearranged into their physical positions in
2D on the sky. Because of the hexagonal close packing arrangement of
the lenslets, alternate rows are offset by half a lens: in making the
cubes this was accounted for by regridding each lenslet by two in the
spatial dimensions (i.e. each lenslet is 2x2 sub-pixels of diameter
0.18 arcsec).

Although the conditions were non-photometric, relative flux
calibration was obtained through observations of the standard star
HIP105206 ($V = 9.46$ mag), taken on a different (photometric) night
at similar airmass. The standard star spectrum was reduced in the same
way as the science data, and the total flux was obtained by summing
$4\times 4$ lenslets (1.5 arcsec diameter). The spectral type of this
star (A1V) meant that it had few stellar features, so normalisation
through division by an appropriate black body (normalised to the {\it
H}-band magnitude) provided a flux calibration, and corrected for the
atmospheric absorption.\footnote{We note that early A-type stars are
characterised by deep Brackett absorption lines, although a suitable
black-body spectrum is adequate to describe the stellar spectrum's
overall shape, and thus to provide a general relative flux
calibration. One potential problem associated with the standard star's
fairly broad Brackett absorption lines is that using it for flux
calibration may introduce artificial emission lines in our galaxy
spectra. Therefore, we masked out the spectral regions affected by the
Brackett (and helium) absorption lines and replaced these by sections
of the appropriate black-body spectrum, before applying our flux
calibration.} The total effective throughput was determined to be
$\sim 8$ per cent on the sky for the middle of the {\it H} band. Since
our observations were taken under non-photometric conditions, this
essentially implies that we can at best obtain lower limits to the
derived fluxes.

\subsection{Imaging}

As part of {\sl HST} programme GO-8645, we obtained observations of NGC 1140
through the F300W (``UV'') and F814W ({\it I$\,$}) filters (Windhorst et al.
2002), with the galaxy centre located on chip 3 of the Wide Field Planetary
Camera 2 (WFPC2). Observations in the F336W ({\it U$\,$}), F555W ({\it V$\,$})
and F785LP (broad {\it I}) passbands, obtained with the Wide Field/Planetary
Camera (WF/PC; PC, chip 5), i.e., prior to the first {\sl HST} refurbishment
mission, were retrieved from the {\sl HST} Data Archive. We have summarised
our combined set of broad-band {\sl HST} observations in Table \ref{obs.tab}.

\begin{table}
\caption[ ]{\label{obs.tab}Overview of the {\sl HST} observations of NGC
1140
}
{\scriptsize
\begin{center}
\begin{tabular}{lrccc}
\hline
\hline
\multicolumn{1}{c}{Filter} & \multicolumn{1}{c}{Exposure time} &
\multicolumn{1}{c}{Centre$^a$} & \multicolumn{1}{c}{PID$^b$} &
\multicolumn{1}{c}{ORIENT$^c$} \\
& \multicolumn{1}{c}{(sec)} & & & \multicolumn{1}{c}{($^\circ$)} \\
\hline
F300W  & 1800             & WF3 & 8645 & 71.93 \\
F336W  & 2$\times$350, 1200, 1400      & PC1 & 2389 & 68.88 \\
F555W  & 2$\times$100, 2$\times$400    & PC1 & 2389 & 68.88 \\
F785LP & 60, 2$\times$230 & PC1 & 2389 & 68.88 \\
F814W  & 200              & WF3 & 8645 & 71.93 \\
\hline
\end{tabular}
\end{center}
{\sc Notes:} $^a$ -- Location of the galactic centre; $^b$ -- {\sl HST}
programme identifier; $^c$ -- Orientation of the images (taken from the
image header), measured North through East with respect to the V3 axis
(i.e., the X=Y diagonal of the WF3 CCD $+ 180^\circ$). 
}
\end{table}

Pipeline image reduction and calibration of the WF/PC and WFPC2 images were
done with standard procedures provided as part of the {\sc
iraf/stsdas}\footnote{The Image Reduction and Analysis Facility (IRAF) is
distributed by the National Optical Astronomy Observatories, which is operated
by the Association of Universities for Research in Astronomy, Inc., under
cooperative agreement with the National Science Foundation. {\sc stsdas}, the
Space Telescope Science Data Analysis System, contains tasks complementary to
the existing {\sc iraf} tasks. We used Version 3.0 (August 2002) for the data
reduction performed in this paper.} package, using the updated and corrected
on-orbit flat fields and related reference files most appropriate for the
observations.

\begin{figure}
\begin{center}
\psfig{figure=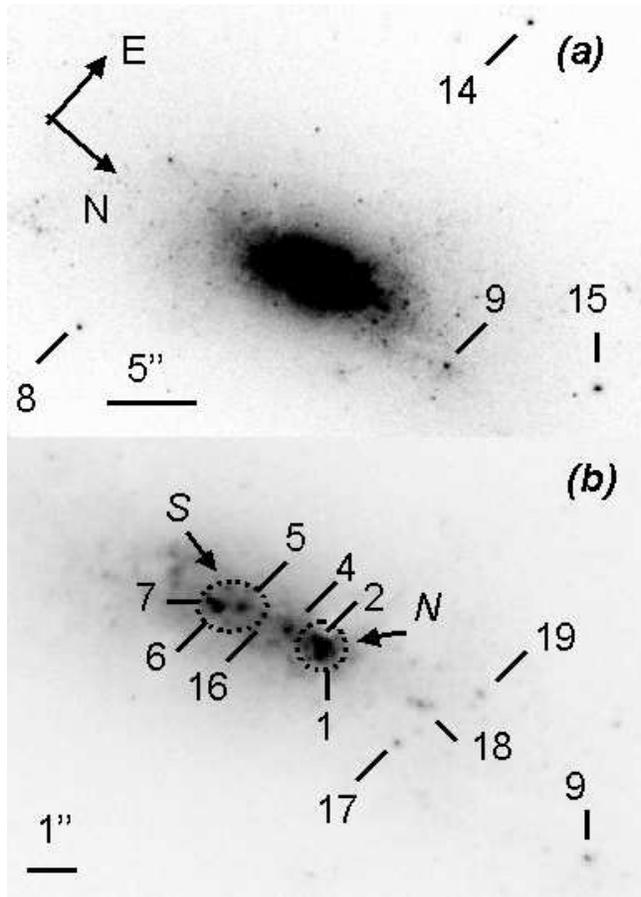,width=8.5cm}
\end{center}
\caption{\label{objects.fig}NGC 1140 young compact cluster
identifications, overlaid on our new WFPC2 F814W images. Panel (a)
shows the general area of the starburst core, with a contrast chosen
to emphasize the large number of fainter compact objects that were not
detected on the older WF/PC images. In panel (b) we show a zoomed-in
section of the very core of the galaxy, with starburst regions S and N
indicated by dotted ellipses. Both panels have the same orientation; 1
arcsec corresponds to 100 pc at the distance of NGC 1140.}
\end{figure}

Our selection of the (compact) YMC sample in NGC 1140's central
starburst is limited by the WF/PC image quality. We therefore start
with the H94a sample of 14 objects, of which objects [H94a]1--9 and 14
are also found on the WFPC2 field of view (FoV; see
Fig. \ref{objects.fig} for the cluster identifications). Careful
comparison of the WFPC2 and WF/PC images reveals five additional
fainter, but well-defined cluster-type objects, subsequently referred
to as objects 15 -- 19. The coordinates of all objects in common on
the WF/PC and WFPC2 FoVs are listed in Table \ref{objects.tab}, and
shown in Fig. \ref{objects.fig}.

\begin{table*}
\caption[ ]{\label{objects.tab}Aperture-corrected {\sc stmag} photometry of
the NGC 1140 objects in common on the WF/PC and WFPC2 FoVs
}
{\scriptsize
\begin{center}
\begin{tabular}{rrrccccc}
\hline
\hline
\multicolumn{1}{c}{ID$^a$} & \multicolumn{1}{c}{R.A. (J2000)} &
\multicolumn{1}{c}{Dec. (J2000)} & \multicolumn{1}{c}{$m_{\rm F300W}$} &
\multicolumn{1}{c}{$m_{\rm F336W}$} & \multicolumn{1}{c}{$m_{\rm F555W}$} &
\multicolumn{1}{c}{$m_{\rm F785LP}$} & \multicolumn{1}{c}{$m_{\rm F814W}$} \\
 & \multicolumn{1}{c}{(h mm ss.sss)} & \multicolumn{1}{c}{(dd mm ss.ss)} &
\multicolumn{1}{c}{(mag)} & \multicolumn{1}{c}{(mag)} & 
\multicolumn{1}{c}{(mag)} & \multicolumn{1}{c}{(mag)} & 
\multicolumn{1}{c}{(mag)} \\
\hline
1+2+3 & 2 54 33.079 &$-$10 01 39.66& 16.26 $\pm$ 0.02 & 16.72 $\pm$ 0.02 & 17.45 $\pm$ 0.02 & 17.33 $\pm$ 0.03 & 18.94 $\pm$ 0.04 \\
   1  &      33.057 &        39.38 & 16.50 $\pm$ 0.05 & 17.07 $\pm$ 0.02 & 17.93 $\pm$ 0.02 & 17.99 $\pm$ 0.03 & 18.51 $\pm$ 0.05 \\
   2  &      33.072 &        39.62 & 16.37 $\pm$ 0.03 & 18.17 $\pm$ 0.04 & 19.25 $\pm$ 0.06 & 19.48 $\pm$ 0.16 & 19.11 $\pm$ 0.08 \\
   4  &      33.075 &        40.63 & 17.65 $\pm$ 0.05 & 19.68 $\pm$ 0.13 & 20.19 $\pm$ 0.08 & 20.13 $\pm$ 0.07 & 19.66 $\pm$ 0.07 \\
   5  &      33.073 &        42.10 & 17.92 $\pm$ 0.05 & 18.90 $\pm$ 0.06 & 20.18 $\pm$ 0.14 & 19.51 $\pm$ 0.15 & 20.03 $\pm$ 0.13 \\
 6+7  &      33.056 &        42.92 & 18.08 $\pm$ 0.09 & 18.91 $\pm$ 0.08 & 18.26 $\pm$ 0.03 & 17.45 $\pm$ 0.02 & 18.17 $\pm$ 0.03 \\
   6  &      33.058 &        42.67 & 18.47 $\pm$ 0.09 & 19.47 $\pm$ 0.07 & 18.76 $\pm$ 0.02 & 17.96 $\pm$ 0.02 & 18.73 $\pm$ 0.03 \\
   7  &      33.069 &        42.95 & 21.71 $\pm$ 1.51 & 20.23 $\pm$ 0.13 & 19.81 $\pm$ 0.06 & 19.03 $\pm$ 0.06 & 19.86 $\pm$ 0.04 \\
   8  &      32.250 &        56.71 & 22.99 $\pm$ 0.23 & 21.34 $\pm$ 0.44 & 21.63 $\pm$ 0.26 & 20.56 $\pm$ 0.09 & 21.30 $\pm$ 0.03 \\
   9  &      32.930 &        29.64 & 20.07 $\pm$ 0.04 & 20.81 $\pm$ 0.18 & 20.95 $\pm$ 0.10 & 20.52 $\pm$ 0.07 & 20.85 $\pm$ 0.03 \\
  14  &      34.758 &        35.87 & 22.43 $\pm$ 0.18 & $\dots$          & $\dots$          & 18.55 $\pm$ 0.02 & 20.30 $\pm$ 0.02 \\
  15  &      33.179 &        18.31 & 20.77 $\pm$ 0.05 & $\dots$          & $\dots$          & $\dots$          & 20.04 $\pm$ 0.02 \\
  16  &      33.072 &        41.38 & 19.76 $\pm$ 0.41 & $\dots$          & $\dots$          & $\dots$          & 21.38 $\pm$ 0.34 \\
  17  &      32.968 &        36.24 & 22.31 $\pm$ 0.52 & 20.40 $\pm$ 0.14 & 21.73 $\pm$ 0.18 & 20.97 $\pm$ 0.14 & 21.69 $\pm$ 0.11 \\
  18  &      33.058 &        36.14 & 20.20 $\pm$ 0.12 & 23.82 $\pm$ 2.76 & 21.88 $\pm$ 0.22 & 20.87 $\pm$ 0.15 & 20.29 $\pm$ 0.05 \\
  19  &      33.127 &        34.66 & 20.12 $\pm$ 0.09 & 20.72 $\pm$ 0.14 & 21.51 $\pm$ 0.15 & 20.73 $\pm$ 0.10 & 21.51 $\pm$ 0.11 \\
\hline
\end{tabular}
\end{center}
}
\flushleft
{\sc Note:} $^a$ Nomenclature of objects 1--9 and 14 is from H94a, see also
Fig. \ref{objects.fig}; objects 15--19 are new additions, from this paper. The
properties of complexes 1+2+3 and 6+7 are based on the integrated fluxes in
these regions, and are therefore based on different aperture sizes than used
for the individual clusters contained within them.
\end{table*}

We obtained {\sc daophot} aperture photometry for all of the objects in Table
\ref{objects.tab}, in all passbands. The correct choice of source and
background aperture sizes is critical for the quality of the resulting
photometry. Due to the varying source separations and background flux levels,
we concluded that we had to assign apertures for source flux and background
level determination individually to each cluster candidate by visual
inspection. Our photometry includes most, if not all, of the light of each
YMC/star-forming region. At the distance of NGC 1140, an aperture of radius 4
-- 7 pixels ($\sim 0.4 - 0.7$ arcsec) corresponds to a projected linear
diameter of $\sim 390 - 680$ pc. Compact YMCs (as well as Galactic GCs) are
characterised by effective radii of $\lesssim 10$ pc (see, e.g., de Grijs et
al. 2001, Whitmore 2003, and references therein), so that such objects would
appear as point sources in our NGC 1140 images. Therefore, we applied
point-source aperture corrections to our photometry, based on comparisons with
artificial PSFs generated using the Tiny Tim software package (Version 6.1, 6
May 2003; Krist \& Hook 2003).

The photometric calibration, i.e., the conversion of the instrumental aperture
magnitudes, thus obtained, to the {\sl HST}-flight system ({\sc stmag}), was
done by simply using the appropriate zero-point offsets for each of the
individual passbands, after correcting the instrumental magnitudes for
geometric distortions, charge transfer (in)efficiency effects, and the
exposure times (see de Grijs et al. [2002] for an outline of the data
reduction procedures). The resulting {\sc stmag} magnitudes for our objects
are also listed in Table \ref{objects.tab}. The uncertainties are the formal
uncertainties, including those due to variations in the background annuli and
Poisson-type noise.

Special care needs to be taken when calibrating the F300W and F336W aperture
magnitudes. These filters suffer from significant ``red leaks'' (Biretta et
al. 2000, chapter 3). Fortunately, the response curve of the F300W red-leak
region resembles the transmission curve of the F814W filter, being most
dominant in the $7000-9000${\AA} range; the red leak of the F336W filter is
roughly similar to the red half of the F814W filter response function. This
implies that we can use our F814W observations to correct, to zeroth order,
the F300W and F336W fluxes for red-leak contamination. For starburst galaxies
dominated by young hot stellar populations the red-leak contamination should
be almost negligible (Biretta et al. 2000). Adopting Eskridge et al.'s (2002)
most straightforward assumption, i.e., that the red leak cannot account for
any more than all of the counts in any given area of a few pixels in the F300W
and/or F336W images, we derive a maximum contribution of the red leak in NGC
1140 to both the F300W and the F336W filters of $\lesssim 1$ per cent of the
total count rates in the respective images. We applied these corrections to
the F300W and F336W images before obtaining the aperture photometry.

\begin{figure*}
\begin{center}
\hspace*{1cm}
\psfig{figure=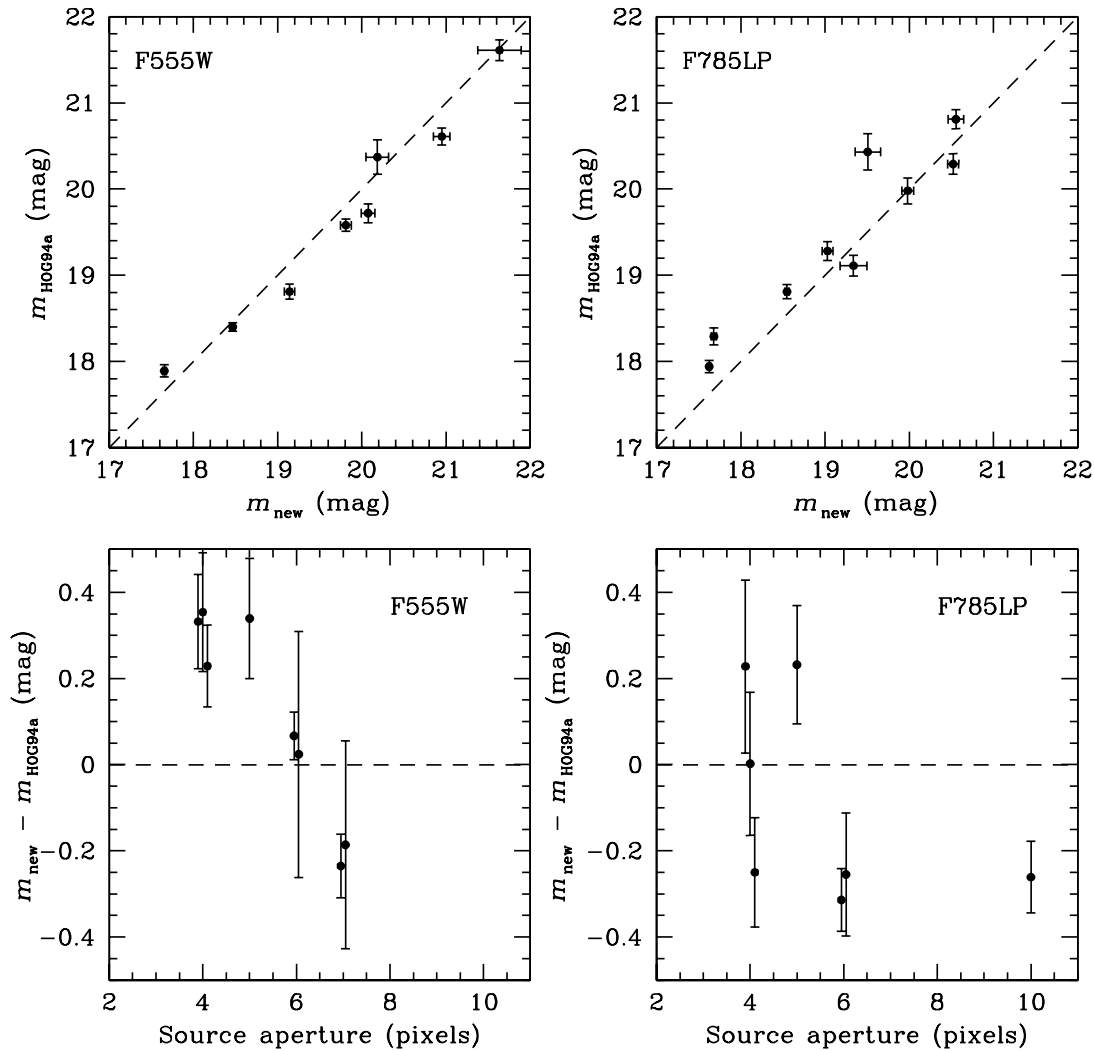,width=15cm}
\end{center}
\caption{\label{compare.fig}Comparison of our aperture-corrected {\sc stmag}
cluster photometry with that of H94a (total fluxes). Equality between both
magnitude measurements is indicated by the dashed lines. The bottom panels
show that there is some dependence, in particular for the F555W passband, of
the magnitude difference on the aperture size used.}
\end{figure*}

In Fig. \ref{compare.fig} we compare our calibrated aperture-corrected
WF/PC photometry with the previously published {\sc stmag} values of
H94a (representing the objects' total fluxes). H94a decided to exclude
the F336W images from their analysis because of (i) the fairly low
exposure levels compared to their longer-wavelength observations, and
(ii) the filter's potentially serious red leak for which they could
not correct. While our F555W and F785LP magnitudes for the objects in
common generally match theirs relatively closely (the
r.m.s. difference in both the F555W and the F785LP filters is 0.22
mag), there is some dependence on the choice of our apertures, in
particular for the F555W magnitudes. The slight dependence of the
relationship between aperture size and magnitude difference implies
either that we may have somewhat underestimated our point-source
aperture corrections compared to the PSF fits of H94a, or that some of
the objects might be marginally extended star-forming regions rather
than individual YMCs (see Section \ref{hst2.sec} for a discussion of
the relevant implications for our results if this were the case), or a
combination of both. Some of the scatter in the data points, however,
is caused by H94a's choice of background annulus (8 -- 15 pixels),
which is not in all cases appropriate in view of the object density in
the galaxy's central starburst region. In our re-analysis of the WF/PC
images we have taken great care to include as much of the source flux
as practically possible without being affected significantly by
neighbouring objects or background fluctuations, and to select
suitable background annuli, on a per-object basis.

\section{CIRPASS spectroscopy of the central starburst}
\label{cirpass.sec}

\begin{figure}[h!]
\begin{center}
\psfig{figure=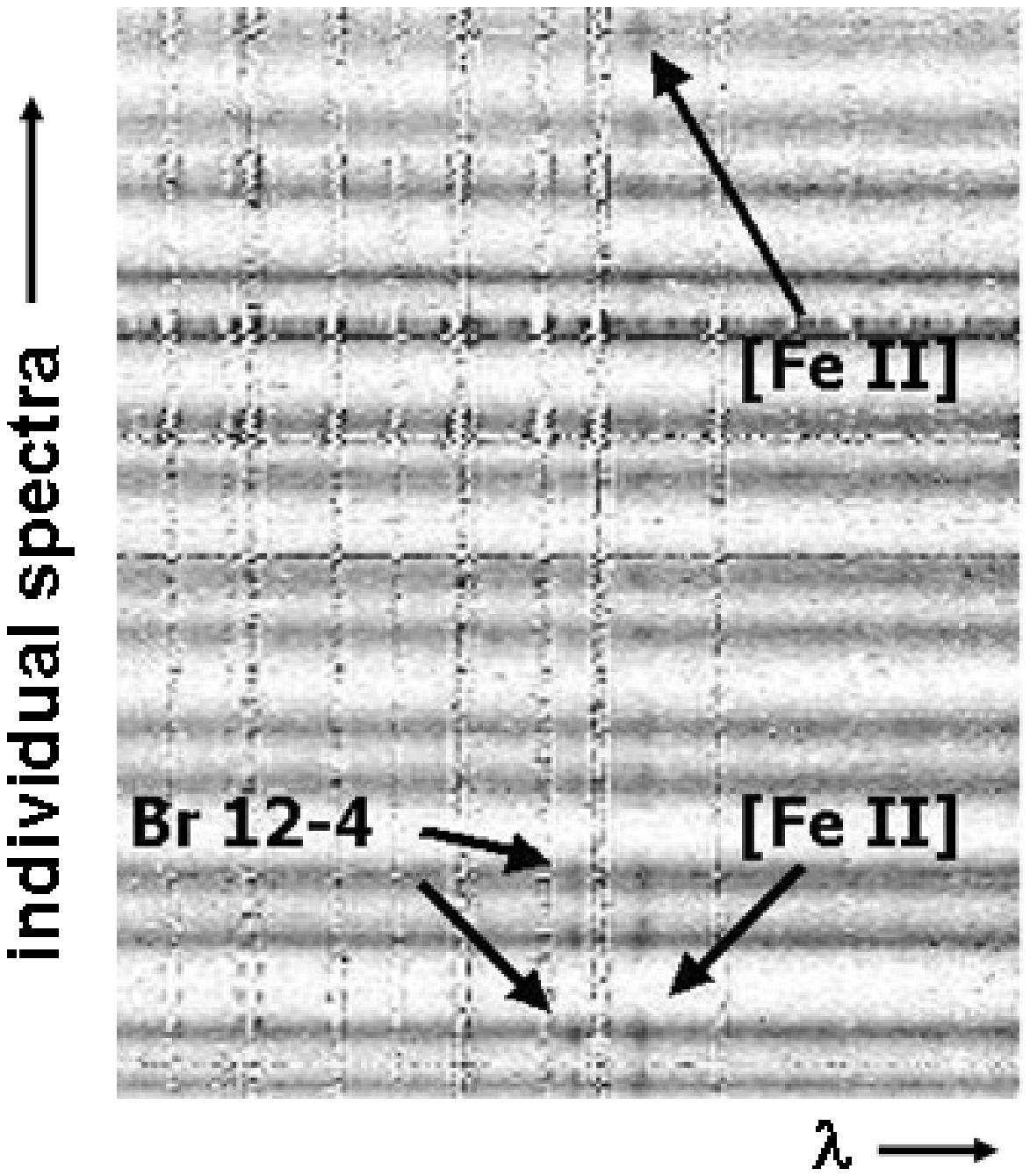,width=8.5cm}
\end{center}
\caption{\label{2d.fig}Section of the 2D reduced set of CIRPASS
spectra, covering the wavelength range from 1.629 to 1.668 $\mu$m, and
displayed using an inverted look-up table to emphasize the emission
lines (dark). Each row corresponds to one of the CIRPASS fibre
spectra, while the horizontal axis carries the wavelength
information. The [Fe {\sc ii}] and Br 12--4 emission lines are
indicated (the Br 13--4 and Br 14--4 emission lines are located
outside of this particular section of the 2D frame); note that while
the [Fe {\sc ii}] emission is found throughout the galaxy (seen as the
dark patches in a large number of the individual spectra), the Br
12--4 emission is only detected in selected spectra. The vertical
features crossing the entire 2D section are sky line residuals, which
could not be fully removed because of the variability of the NIR sky
background; this section of the 2D spectrum represents our best data
reduction effort (as can be seen from the very flat continuum in
between the sky lines). It shows the difficulty of working with NIR
IFU spectroscopy. See Section \ref{spectroscopy.sec} for a full
technical description.}
\end{figure}

\begin{figure*}
\begin{center}
\hspace*{0.1cm}
\psfig{figure=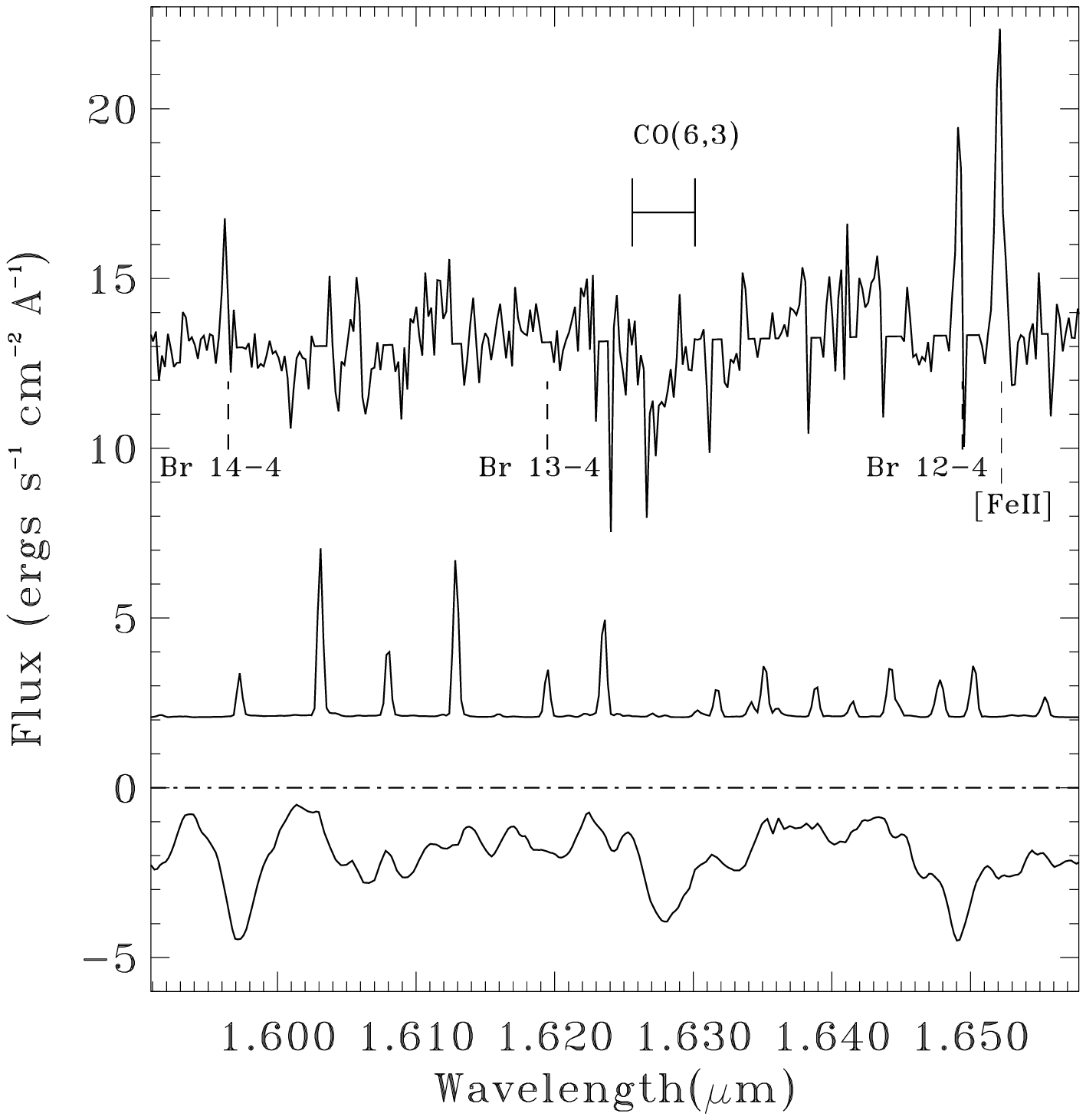,width=8.5cm}\psfig{figure=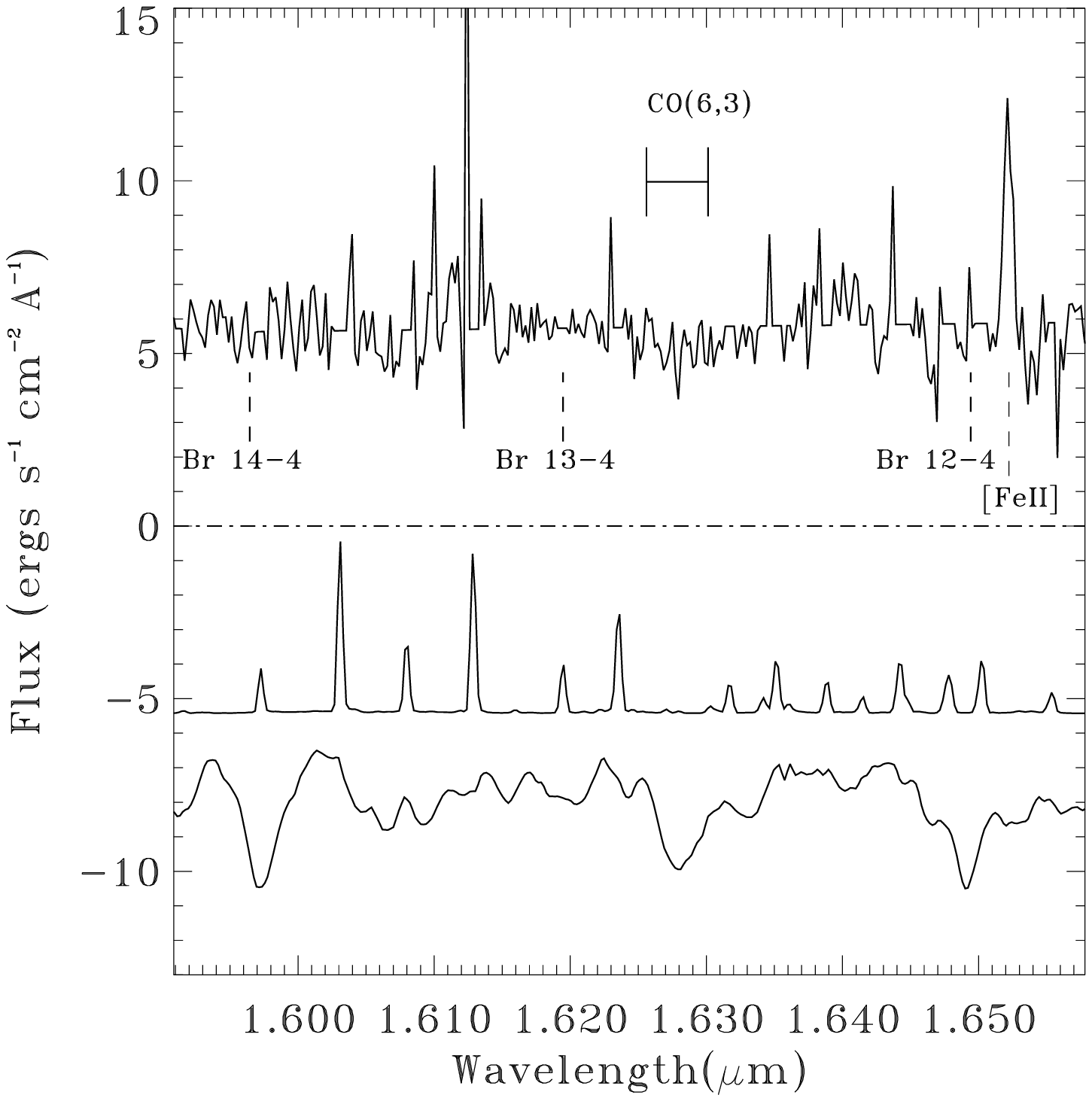,width=8.5cm}
\end{center}
\caption{\label{spectra.fig}Integrated spectra of regions NGC 1140-N
(left) and NGC 1140-S (right), as described in the text. The positions
of the sky lines, shown as the middle spectra, have been masked out --
although residuals remain visible (cf. Fig. \ref{2d.fig}); the
redshifted wavelengths expected for the [Fe {\sc ii}] and the
high-order Brackett emission lines, and also for the CO(6,3)
absorption feature are indicated. The bottom spectra are (arbitrarily
scaled) theoretical spectra of simple stellar populations (SSPs) at an
age of 10 Myr, based on the Starburst99 SSP models (Leitherer et
al. 1999), which have better spectral resolution at these NIR
wavelengths than the Anders \& Fritze-v. Alvensleben (2003) SSP models
that we prefer to use for our broad-band spectral energy distribution
fits at optical wavelengths (Section \ref{hst1.sec}).}
\end{figure*}

The wavelength range from $\sim 1.45 - 1.67 \mu$m covers a number of
emission lines expected to be strong in star-forming areas and active
starburst regions. Despite the strong OH emission lines affecting this
section of the {\it H} band (with origins in the Earth's atmosphere),
we identified four strong emission lines in a significant fraction of
the lenslets covering the NGC 1140 starburst, three of which were
located in spectral regions well away from the sky lines, and clearly
resolved. This is largely thanks to the relatively high spectral
resolution of the CIRPASS instrument. In addition to strong [Fe {\sc
ii}] emission (at $\lambda_{\rm rest} = 1.644 \, \mu$m), we also
detect three high-order Brackett (hydrogen) lines, Br 12--4
($\lambda_{\rm rest} = 1.641 \, \mu$m), Br 13--4 ($\lambda_{\rm rest}
= 1.611 \, \mu$m) and Br 14--4 ($\lambda_{\rm rest} = 1.588 \,
\mu$m). However, the Br 13--4 flux is significantly compromised by a
nearby strong sky line. In Fig.  \ref{2d.fig} we show a small section
of our final, reduced set of spectra, including the [Fe {\sc ii}] and
Br 12--4 lines, using an inverted look-up table to emphasize the
emission lines; in Fig. \ref{spectra.fig} we show the integrated
spectra of both starburst regions in a more traditional way, combining
24 and 17 individual CIRPASS lenses for the spectra of regions NGC
1140-N and S, respectively. The [Fe {\sc ii}] line on the one hand,
and the high-order Brackett lines on the other arise through different
physical mechanisms (see Section \ref{context1.sec} for a full
discussion). We also detect a weak absorption feature at the
wavelength range expected for the CO(6,3) band head (at $\lambda
\simeq 1.627 \, \mu$m, after correcting for the galaxy's redshift), in
the most active starburst regions, in particular in NGC 1140-N. In
this northern starburst region, we measure an equivalent width (EW) of
the CO feature of $\sim 5${\AA} (using the Origlia, Moorwood \& Oliva
[1993] models as our template), which we consider a reasonably robust
measurement despite the poor signal-to-noise (S/N) ratio. In the
southern starburst region, the continuum was detected but the S/N
ratio was not sufficient for reliable absorption-line studies, or EW
measurements. In addition, there is a lack of theoretical analyses of
the {\it H}-band CO feature at the spectral resolution offered by
CIRPASS. This therefore presents a good case for follow-up
spectroscopy in the {\it K} band, where the CO features are much
stronger.

\begin{figure}
\begin{center}
\psfig{figure=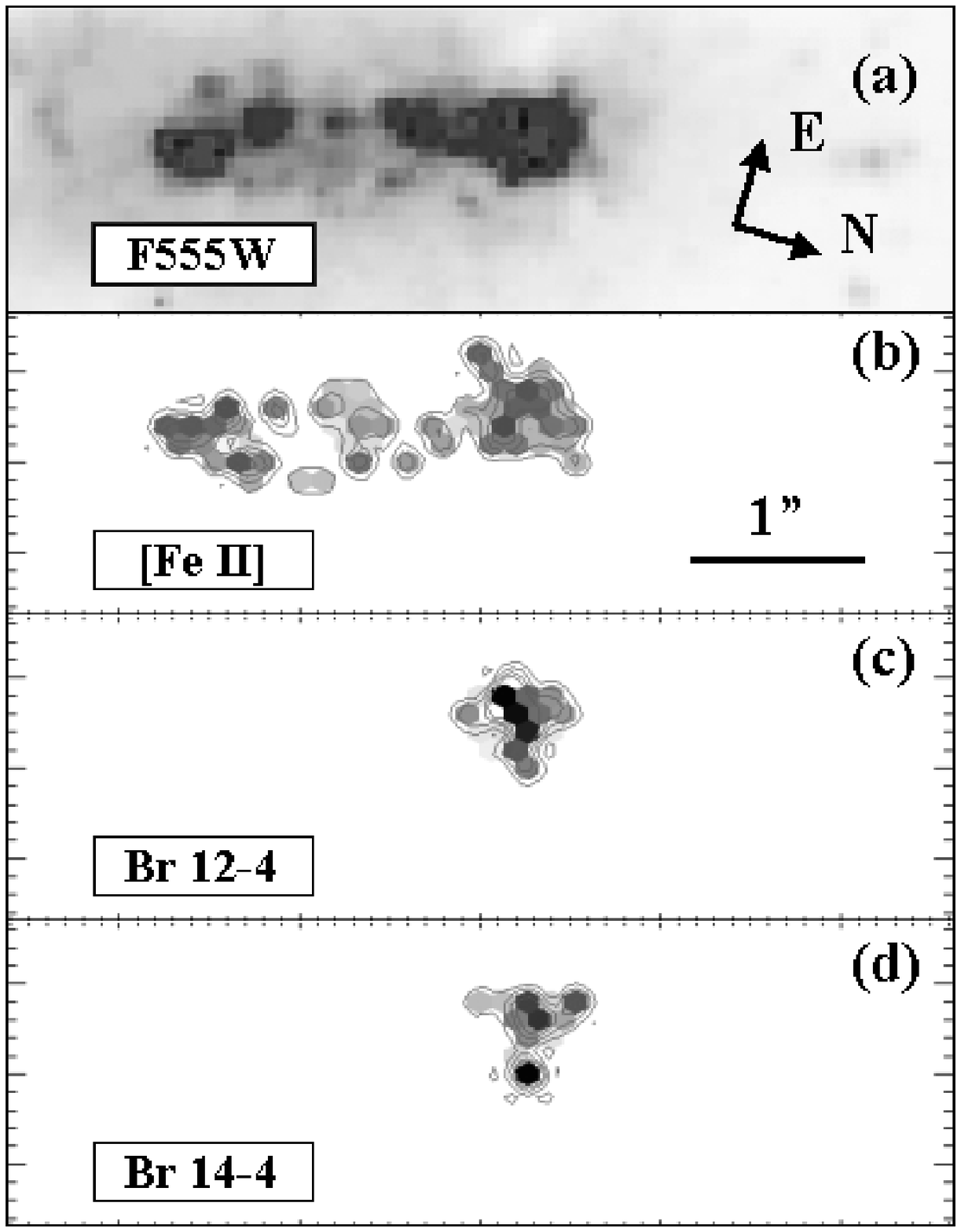,width=8.5cm}
\end{center}
\caption{\label{emission.fig}Maps of (a) the {\sl HST}/WFPC2 F555W broad-band
continuum flux distribution, (b) the [Fe {\sc ii}] emission, (c) the Br 12--4,
and (d) the Br 14--4 emission-line morphologies of NGC 1140. The CIRPASS data
show both the flux levels in the individual fibre lenses (above $\sim 3
\sigma$, where $\sigma$ is the noise in the continuum level of the spectra
near the respective emission lines), as well as smoothed contours of the same
data, in order to facilitate large-scale comparisons of the morphologies. The
Br 12--4 and Br 14--4 fluxes originate entirely in region NGC 1140-N; region
NGC 1140-S corresponds to the southernmost 1 arcsec region with strong [Fe
{\sc ii}] emission. The peak fluxes in NGC 1140-N in the [Fe {\sc ii}], Br
12--4 and Br 14--4 emission line maps are 1.1, 1.2, and $1.2 \times 10^{-17}$
erg s$^{-1}$ cm$^{-2}$ {\AA}$^{-1}$, respectively, with the continuum level
being close to $0.9 \times 10^{-17}$ erg s$^{-1}$ cm$^{-2}$ {\AA}$^{-1}$ in
all cases. The [Fe {\sc ii}] peak flux of NGC 1140-S is $0.8 \times 10^{-17}$
erg s$^{-1}$ cm$^{-2}$ {\AA}$^{-1}$ (with a continuum level of $0.3 \times
10^{-17}$ erg s$^{-1}$ cm$^{-2}$ {\AA}$^{-1}$). The likely (realistic)
uncertainties are on the order of 30 per cent in all cases.}
\end{figure}

In Fig. \ref{emission.fig} we show the flux distribution of the [Fe
{\sc ii}], Br 12--4 and Br 14--4 emission lines across the face of NGC
1140, compared to the higher resolution {\sl HST} F814W
morphology. The flux distribution of the [Fe {\sc ii}] line is clearly
more extended than those of the Brackett lines.  While [Fe {\sc ii}]
emission is found throughout the galaxy, the emission from all of the
high-order Brackett lines (including Br 12--4, Br 13--4 and Br
14--4) is found predominantly associated with the northern starburst
region, including the compact YMCs [H94a]1--3. We will refer to the
southern and northern starburst areas, traced in turn by the [Fe {\sc
ii}] emission and by both the [Fe {\sc ii}] and Br lines, as NGC
1140-S and N, respectively. These two starburst regions include
clusters [H94a]5--7 and [H94a]1--3, respectively. Our flux
measurements -- and other spectral features -- associated with these
regions are based on the composite properties of the 17 and 24
individual CIRPASS lenses centred on the peak of the [Fe {\sc ii}]
emission in regions S and N, respectively. The morphologies of the
lines' EWs and FWHMs show a similar, and consistent, picture; the
largest EWs and FWHMs are found coincident with the highest
intensities (see Table \ref{fluxes.tab} for the relevant flux and EW
measurements).  Although the observing conditions under which the
CIRPASS data were obtained were non-photometric, the fact that we
observe all wavelengths simultaneously over the entire 2D FoV enables
us to say with confidence that this different spatial distribution is
real. This is one of the major advantages of using IFUs.

By (median) combining the [Fe {\sc ii}] flux in the starburst regions
NGC 1140-S and N, we find $\lambda_{\rm central} {\rm ([Fe {\sc ii}])}
= 16521.6$ and 16520.5{\AA}, respectively, and respective composite
EWs for these regions of EW$_{\rm [Fe {\sc ii}]} = 8$ and 4{\AA}; the
uncertainties in the line centres and line widths are both $\sim
0.4${\AA}, while the uncertainties in the EW measurements are on the
order of 30 per cent. The mean of the [Fe {\sc ii}] line centres
corresponds to a NIR redshift for the galaxy of $z \simeq 0.00493\;
(\pm 3 \times 10^{-5})$ or, equivalently, a heliocentric radial
velocity of $v_\odot \simeq 1480 \pm 9$ km s$^{-1}$ (where the
uncertainty includes both the difference in the line centres between
the northern and southern starburst regions, and also the uncertainty
in the actual line centre mesurements), which is in the same range as
the most recent optical redshift measurements (e.g., $z = 0.0048$,
Dahari 1985; $v_\odot = 1498 \pm 33$ km s$^{-1}$, RC3), and slightly
lower than recent measurements based on H{\sc i} 21cm data (e.g.,
$v_\odot = 1509 \pm 4$ km s$^{-1}$, RC3; $v_\odot = 1501 \pm 1$ km
s$^{-1}$, Haynes et al. 1998). The composite Br 12--4 and Br 14--4
lines in NGC 1140-N are centred on $\lambda_{\rm central}$ (Br 12--4)
= 16491.5{\AA} and $\lambda_{\rm central}$ (Br 14--4) = 15962.0{\AA},
respectively. This corresponds to redshift measurements of $z =
0.00497$ and $z = 0.00516$ for Br 12--4 and Br 14--4 in this region,
respectively. The maximum offset associated with bulk peculiar motions
in the gas is $<40$ km s$^{-1}$.

The overall Br 14--4 to Br 12--4 flux ratio in the most active
extended starburst region, NGC 1140-N, is $F($Br 14--4$) / F($Br
12--4$) = (13 \times 10^{-17}) / (28 \times 10^{-17}) \simeq 0.5$
(where the individual flux measurements are in units of erg s$^{-1}
{\rm cm}^{-2}$; see Table \ref{fluxes.tab}), with an uncertainty of
roughly 50 per cent because of the low S/N ratio of both emission
lines outside the very core of region N, in particular for Br 14--4;
the likely total systematic uncertainties in the individual
measurements (caused by both calibration and random uncertainties) are
on the order of 30 per cent. This line ratio is, within the
uncertainties, comparable to the theoretical predictions of Hummer \&
Storey (1987) for Case B recombination physics, who -- for a range of
electron densities and temperatures -- calculate line ratios upward of
0.6 for these transitions.

\section{Star cluster properties revealed by the {\sl Hubble Space Telescope}}
\label{hst.sec}

\subsection{Technique and main results}
\label{hst1.sec}

In Anders et al. (2004a; see applications in Anders et al. [2004b] and
de Grijs et al. [2003c]) we developed a reliable method to determine
cluster ages (in the range from a few $\times 10^7$ to a few $\times
10^{10}$ yr), masses, metallicities and extinction values robustly and
simultaneously based on high-resolution {\sl HST}/WFPC2 imaging
observations in a minimum of four broad-band passbands covering the
entire optical wavelength range from the UV/{\it U} to the {\it I}
band (and preferably extending to the NIR) or their equivalents in
non-standard passband systems (see also de Grijs et al. [2003a], in
particular for a discussion on the systematic uncertainties involved
in estimating the individual YMC metallicities). We tested our method
using both artificial star clusters of various ages (Anders et
al. 2004a) and also by using the $\sim 150$ young clusters in the
centre of the nearby starburst galaxy NGC 3310, and confirmed the
previously suggested starburst scenario in that galaxy (de Grijs et
al. 2003c). By combining our new WFPC2 observations with the archival
WF/PC data, we span at least as long a wavelength range with adequate
passband separation, although the WF/PC spatial resolution is
significantly poorer than that of WFPC2.

With the lessons learnt from Anders et al. (2004a) and de Grijs et al.
(2003a,c) in mind, in particular regarding the systematic
uncertainties involved in using broad-band SEDs to obtain individual
YMC properties, we applied a similar three-dimensional (3D) $\chi^2$
minimisation to the SEDs of our NGC 1140 YMCs and star-forming
complexes. The minimisation was done with respect to the Anders \&
Fritze--v. Alvensleben (2003) simple stellar population (SSP) models,
which include the contributions of an exhaustive set of nebular
emission lines and gaseous continuum emission, shown to be important
during the first few $\times 10^7$ yr, depending on the metallicity
assumed. We thus obtained, for each individual object, the most likely
combination of age, mass, and internal extinction, E$(B-V)_{\rm
int}$. For our mass estimates we assumed a Salpeter-type IMF with a
lower-mass cut-off around 0.08 M$_\odot$ and an upper mass cut-off
between 50 and 70 M$_\odot$, the exact mass depending on the cluster's
metallicity, as provided by the Padova isochrones used for our SSP
comparison (see in particular Schulz et al. 2002 and Anders et
al. 2004a).

We realise that recent determinations of the stellar IMF deviate
significantly from a Salpeter-type IMF at low masses, in the sense
that the low-mass stellar IMF is significantly flatter than the
Salpeter slope.  The implication of using a Salpeter-type IMF for our
cluster mass determinations is therefore that we have {\it
overestimated} the individual cluster masses (although the relative
mass distribution of our entire cluster sample remains unaffected).
Therefore, we used the more modern IMF of Kroupa, Tout \& Gilmore
(1993) to determine the correction factor, $C$, between our masses and
the more realistic masses obtained from the Kroupa et al. (1993) IMF
(both normalised at $1.0 M_\odot$).  This IMF is characterised by
slopes of $\alpha = -2.7$ for $m > 1.0 M_\odot$, $\alpha = -2.2$ for
$0.5 \le m/M_\odot \le 1.0$, and $-1.85 < \alpha < -0.70$ for $0.08 <
m/M_\odot \le 0.5$.  Depending on the adopted slope for the lowest
mass range, we have therefore overestimated our individual cluster
masses by a factor of $1.7 < C < 3.5$ for an IMF containing stellar
masses in the range $0.08 \le m/M_\odot \le 70$.

The internal extinction was modelled assuming a Calzetti et al. (2000)
starburst galaxy-type extinction law\footnote{We realise that an
LMC-type extinction law is more appropriate for compact cluster-like
objects in starburst galaxies than the Calzetti et al. (2000) law,
which applies to more extended starburst regions, but note that the
differences between both laws are smaller than the combined
observational and model uncertainties over the wavelength range
covered by our {\sl HST} observations.}; we adopted an extinction
resolution of $\Delta$E$(B-V)_{\rm int} = 0.05$ mag. Galactic
foreground extinction toward NGC 1140 is estimated at $A_{V,{\rm Gal}}
= 0.124$ mag, or E$(B-V)_{\rm Gal} = 0.038$ mag (Schlegel, Finkbeiner
\& Davis 1998), and was corrected for using a standard Galactic
extinction law.

Independently measured metallicities {\it for the compact NGC 1140
YMCs} have not been published, but recent metallicity estimates for
the galaxy's ISM, based on oxygen abundances, $12 + \log({\rm O/H})$,
range from $\sim 8.0$ (Stasi\'nska, Comte \& Vigroux 1986, Marconi,
Matteucci \& Tosi 1994, Heckman et al. 1998) to $\sim 8.5$
(Storchi-Bergmann et al. 1994, Calzetti 1997, Guseva, Izotov \& Thuan
2000). With a solar oxygen abundance of $12 + \log({\rm O/H})_\odot =
8.69 \pm 0.05$ (Allende Prieto, Lambert \& Asplund 2001), this
translates to $Z_{\rm NGC 1140} \sim (0.2 - 0.6) Z_\odot$.

Because of (i) the very significant effects in UV and optical
passbands of the age--metallicity degeneracy for the very young ages
expected (de Grijs et al.  2003a,c), (ii) the coincidence of the
youngest NGC 1140 YMC ages with the youngest age range covered by our
models (Schulz et al. 2002, Anders \& Fritze--v. Alvensleben 2003),
where the contributions of the red supergiant (RSG) stars (which
significantly affect the evolutionary tracks around 10 Myr) are not
yet well-understood (see, e.g., Massey \& Olsen 2003; we use the
Padova tracks in our models; they do include a description of the RSG
phase, which is however very uncertain), (iii) the partial overlap
between the F785LP and F814W filter transmission curves, thus reducing
the robustness with which we can obtain our parameters (Anders et al.
2004a, de Grijs et al. 2003c), and (iv) the lack of the NIR passbands
that we would have used ideally to constrain the YMC metallicities
(Anders et al.  2004a, Parmentier, de Grijs \& Gilmore 2003), we
decided to assume a generic metallicity of $0.4 Z_\odot$ for all of
the compact NGC 1140 YMCs, instead of trying to fit their
metallicities independently. If (i) the starburst event responsible
for the formation of the NGC 1140 YMCs indeed induced (compact) YMC
formation on a galaxy-wide scale on short time-scales, and would
thereby simultaneously cause significant mixing of the ISM (due to,
e.g., ram-pressure effects), and (ii) prior to the recent starburst
event star formation occurred throughout the galaxy at low levels with
roughly comparable star formation rates (averaged over long,
relatively quiescent periods), one should not expect the YMCs to have
significantly varying metallicities. This assumption seems therefore
justified by physical arguments on the one hand, and technical
limitations on the other.

The results of our fits to the ages, masses and (internal) extinction
values are tabulated in Table \ref{basics.tab}, for those YMCs for
which we obtained fits with reasonable minimum $\chi^2$ values. We
show both our best-fit estimates and the most likely parameter ranges
(denoted by `min.' and `max.'), following the technique described in
Anders et al. (2004a) and successfully applied in Anders et
al. (2004b) and de Grijs et al. (2003a,c). The uncertainties include
the range encompassed by the best-fit 31.74 per cent (i.e., $1
\sigma$) of our SSP models around the mean (see Anders et al.  2004a).

\begin{table*}
\caption[ ]{\label{basics.tab}Properties of the compact NGC 1140 YMCs derived
from our multi-passband {\sl HST} photometry, assuming a fixed metallicity of
$0.4 Z_\odot$. Tabulated are our best estimates and the range from the minimum
to the maximum likely values.
}
{\scriptsize
\begin{center}
\begin{tabular}{rccccccrrr}
\hline
\hline
\multicolumn{1}{c}{ID} & \multicolumn{3}{c}{E$(B-V)$ (mag)} &
\multicolumn{3}{c}{Age ($\times 10^7$ yr)$^a$} & \multicolumn{3}{c}{Mass
($\times 10^6 {\rm M}_\odot$)} \\
 & \multicolumn{1}{c}{Min.} & \multicolumn{1}{c}{Best} &
 \multicolumn{1}{c}{Max.} & \multicolumn{1}{c}{Min.} &
 \multicolumn{1}{c}{Best} & \multicolumn{1}{c}{Max.} &
 \multicolumn{1}{c}{Min.} & \multicolumn{1}{c}{Best} &
 \multicolumn{1}{c}{Max.} \\ 
 \hline 
   2  & 0.14 & 0.20 & 0.26 & 1.1 & 1.6 & 2.1 & 0.9 & 1.3 & 1.7 \\ 
   4  & 0.45 & 0.45 & 0.50 & 1.6 & 1.6 & 1.6 & 1.4 & 1.4 & 1.6 \\ 
   5  & 0.00 & 0.00 & 0.00 & 2.4 & 2.4 & 2.8 & 0.4 & 0.4 & 0.5 \\
   6  & 0.50 & 0.50 & 0.60 & 1.2 & 2.0 & 2.0 & 6.3 & 7.1 & 7.5 \\ 
   7  & 0.75 & 0.75 & 0.80 & 1.2 & 1.2 & 1.2 & 3.7 & 3.7 & 4.4 \\ 
   9  & 0.25 & 0.30 & 0.35 & 2.0 & 2.0 & 3.2 & 0.5 & 0.5 & 0.7 \\ 
  17  & 0.30 & 0.35 & 0.35 & 1.2 & 1.2 & 1.2 & 0.2 & 0.2 & 0.2 \\ 
  19  & 0.45 & 0.50 & 0.50 & 1.2 & 1.2 & 1.2 & 0.4 & 0.4 & 0.4 \\ 
\hline
\end{tabular}
\end{center}
}
{\sc Note:} $^a$ -- These ages are derived from fits to broad-band SEDs, and
may have a scale uncertainty toward high ages, as discussed in the text.
\end{table*}

\subsection{Implications}
\label{hst2.sec}

\begin{figure}
\begin{center}
\psfig{figure=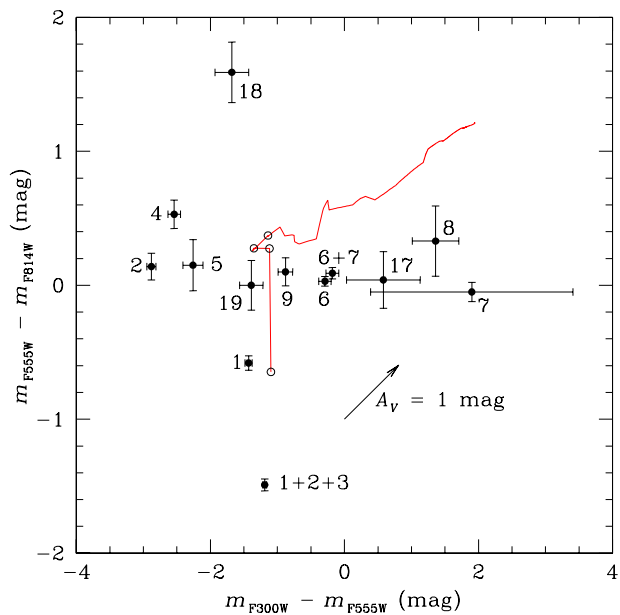,width=8.5cm}
\end{center}
\caption{\label{ccdiag.fig}Distribution of the compact NGC 1140 YMCs
in the two-colour $(m_{\rm F300W} - m_{\rm F555W})$ vs. $(m_{\rm
F555W} - m_{\rm F814W})$ plane. Overplotted is the $0.4 Z_\odot$ model
of Anders \& Fritze--v.  Alvensleben (2003); the open circles on this
model indicate ages of 4, 8, 12, and 20 Myr, respectively, from the
bottom upward. The arrow shows the effect of 1 mag of visual
extinction. The error bars represent the formal uncertainties caused
by pixel noise effects; they do not include the uncertainties owing to
crowding effects and their associated flux contamination.}
\end{figure}

We will illustrate the uncertainties involved in the multi-parameter
analysis of the broad-band SEDs by means of a two-colour diagnostic
figure, Fig.  \ref{ccdiag.fig}. Here, we show the distribution of the
compact NGC 1140 YMCs in the plane defined by the $(m_{\rm F300W} -
m_{\rm F555W})$ and $(m_{\rm F555W} - m_{\rm F814W})$ colours.
Overplotted on the distribution of data points is the $0.4 Z_\odot$
model of the Anders \& Fritze--v. Alvensleben (2003) set of SSPs; the
open circles on this model indicate ages of 4, 8, 12, and 20 Myr,
respectively, from the bottom upward. We also show the effect of 1 mag
of visual extinction. The inclusion of a more dominant RSG phase than
in the present model set would introduce a ``loop'' in the model close
to the area where most of the clusters are concentrated.  In
principle, if our CO band head detection had been of sufficient S/N
ratio and spatial resolution, the presence or absence of this CO
feature in the CIRPASS spectra of the individual YMCs could
potentially provide additional constraints on the importance of the
uncertain RSG phase, although CO features are in fact present at a
relatively large range of ages; showing that they are caused by the
presence of RSGs requires both very good S/N ratios and better models
than currently available.  Unfortunately, the spatial resolution of
our CIRPASS observations is significantly poorer than the {\sl HST}
WFPC2 resolution, and as such we can only conclude that the strongest
starburst region, encompassing clusters [H94a]1,2,4, shows some
evidence for the presence of the CO(6,3) band head feature, although
we do not have enough information to match these low-S/N detections to
any of the individual YMCs.

From this simple diagnostic figure, it follows that our
multi-parameter analysis (Table \ref{basics.tab}) may have resulted in
overestimates of a number of the YMC ages, and that at least the
scatter of data points on the redward side of the model may be
explained by extinction, with a similar degree of confidence. While
intrinsic variations in the cluster IMFs may introduce additional
scatter, such a scenario is not supported by our observations;
high-resolution spectroscopy would be required to resolve this issue
satisfactorily (e.g., Smith \& Gallagher 2001, Mengel et al. 2002). On
the other hand, since we have likely overestimated our YMC masses by a
factor of a few to up to an order of magnitude (e.g., by assuming a
Salpeter IMF, as discussed above, and by overestimating the cluster
ages based on broad-band photometry, see below), stochastic effects
may be responsible for at least part of the spread in
Fig. \ref{ccdiag.fig}: at these masses, of a few $\times 10^4 - 10^5$
M$_\odot$ (corrected for our overestimates), IMF sampling effects
become noticeable and significant (Lan\c{c}on \& Mouhcine 2000,
Bruzual 2002, Bruzual \& Charlot 2003).

Despite the large uncertainties inherent to the broad-band SED
analysis, by combining the diagnostic diagram of Fig \ref{ccdiag.fig}
with the results summarised in Table \ref{basics.tab} we see that
complex [H94a]1--3 may be somewhat younger than complex [H94a]6--7,
despite our relatively coarse age resolution of 4 Myr at these young
ages (with a minimum model age of 4 Myr). [Regarding the individual
YMCs, we could not obtain a satisfactory best-fit result for cluster 1
due to intrinsic problems related to the crowding of its immediate
environment and the associated contamination of the WF/PC fluxes;
cluster 3 could not be identified robustly as a separate compact
object in the WFPC2 images.]

It appears, therefore, that there may be a small age gradient from
complex [H94a]1--3 to complex [H94a]6--7.  Although we justified our
assumption of a mean YMC metallicity of $0.4 Z_\odot$ in Section
\ref{hst1.sec}, we also performed our multi-parameter fits using the
two adjacent steps in metallicity included in our models, i.e., 0.2
and $1.0 Z_\odot$. While -- as expected -- we notice the effects of
the well-known age-metallicity degeneracy, the relative results,
including the hint of an age gradient across the galaxy, remain the
same. In other words, this possible age gradient is unlikely to be
caused by our assumption of the YMC's metallicity, unless there are
significant metallicity variations from cluster to cluster. We will
use these results for our interpretation of the galaxy's recent
star-formation history in Section \ref{context1.sec}. One should keep
in mind, however, that our {\it absolute} age estimates are merely
indicative, because our SSP models are not fully adequate to deal very
robustly with very young star clusters: our lower age limit of 4 Myr
and the 4-Myr age resolution, in addition to the inclusion of
uncertain descriptions in the models, or absence, of stellar
evolutionary phases such as the RSG and Wolf-Rayet stages,
respectively, which play an important role at these ages, render
absolute age estimates rather uncertain (see, e.g., Homeier, Gallagher
\& Pasquali [2002] for the implications of including an uncertain
description of the RSG phase in the models, although Whitmore \& Zhang
[2002] make a case for less significant effects, based on their
observations of YMCs in the Antennae galaxies). These effects will
tend to make us overestimate the ages of the youngest objects, with
ages within our youngest two age bins. However, relative age
estimates, and therefore age gradients, are much more robust and can
more reliably be used as diagnostics (see de Grijs et al. 2003a,c,
Anders et al. 2004a), in particular at these young ages, where the
colour evolution is rapid.

Our analysis in this paper represents the first one-to-one comparison
of medium-resolution spectroscopy results with results from fits to
broad-band {\sl HST} SEDs using SSP models with newly updated input
physics, most importantly the inclusion of an extensive set of gaseous
emission-line and continuum emission. The large uncertainties
resulting from the broad-band SEDs fits compared to the more accurate
CIRPASS results therefore serve as an important lesson regarding the
reliability of imaging data in the context of SSP evolutionary
scenarios.

A conclusive test of the proposed scenario giving rise to an age
gradient across the galaxy would be to obtain narrow-band He{\sc
ii}$\lambda 4686$ observations of the galaxy, a prime diagnostic of
the presence of Wolf-Rayet stars associated with active starburst
events. If one can obtain observations of high spatial resolution, so
that the YMCs can be well separated from unrelated contamination from
nearby objects, additional high-S/N Balmer-line EWs could potentially
also provide evidence for the suggested age gradient.

Our photometric mass estimates of the compact NGC 1140 YMCs are
comparable to the Galactic GCs comprising the high-mass wing of the GC
mass distribution, even if we have overestimated our cluster masses by
up to a factor of 3.5 (e.g. Mandushev, Staneva \& Spassova 1991, Pryor
\& Meylan 1993) and to the spectroscopically confirmed masses of the
so-called ``super star clusters'' in M82 (M82-F; Smith \& Gallagher
2001), and the Antennae galaxies (Mengel et al.  2002). Our detection
of similarly massive YMCs in NGC 1140 supports the scenario that such
objects form preferentially in the extreme environments of interacting
and starburst galaxies (cf. de Grijs et al. 2001, 2003a). This galaxy
adds to the number of cases where rapid star formation has evidently
produced compact YMCs. A subsequent paper (R. de Grijs et al., in
prep.) will be dealing with the photometric versus spectroscopic mass
estimates of the YMCs in NGC 1140. If we overestimated the ages of the
brightest objects significantly, due to the model limitations
discussed above, the actual masses of these objects will be lower, by
an additional factor of $\sim 2-4$, somewhat depending on the
metallicity. As a caveat inherent to the nature of our data and its
analysis, it should be realised, however, that even at {\sl HST}
resolution, point-like sources at the distance of NGC 1140 may in fact
be more extended star-forming regions or multiple subclusters (and
therefore potentially physically different from GC-type progenitors),
although the vast majority of the compact YMC candidates in NGC 1140
have effective sizes smaller than the {\sl HST} PSF.

H94a assumed a generic internal extinction of $A_{\rm F555W} = 0.3$ mag toward
all YMCs in NGC 1140, based on the galaxy's Balmer decrement (H94b). The
average internal extinction in NGC 1140 appears to be moderately low overall,
with estimates ranging from E$(B-V) = 0.10 - 0.18$ mag (Storchi-Bergmann et
al. 1994, H94b, Calzetti et al. 1994, 1995, Calzetti 1997), based on detailed
modelling of the internal dust properties, or $A_V \sim 0.33 - 0.59$ mag
(assuming a Milky Way-type extinction law). This is consistent with the low
extinction in the H$\alpha$ emission line measured by Buat et al. (2002),
$A_{{\rm H}\alpha} = 0.25$ mag, and with the low $V$-band effective optical
depths of $\tau_V^{\rm eff} \sim (0.4 - 0.8)$ derived by Takagi et al. (1999,
2003; but see Section \ref{context2.sec}). The latter results are fairly
independent of both the extinction law assumed and the mass fraction of the
galaxy involved in the starburst. Rosa-Gonz\'alez, Terlevich \& Terlevich
(2002), on the other hand, derive a significantly higher global internal
extinction, of $A_V \sim (0.62 - 0.93)$ mag. Nevertheless, it should be kept
in mind that a fairly low global extinction estimate does not preclude a
patchy extinction distribution. In addition, most of these extinction
estimates are based on H$\alpha$/H$\beta$ line flux ratios, which introduces a
bias towards the more actively star-forming regions. Thus, our varying
extinction estimates for the YMCs in the central starburst (ranging between
$A_V \simeq 0$ and $A_V \simeq 2.6$ mag, with most clusters being affected by
$A_V \lesssim 1$ mag; cf. Fig. \ref{ccdiag.fig}) are not inconsistent with the
low extinction estimates derived for the galaxy as a whole. Based on a
comparison of our F300W/F336W and F814W images, it appears indeed that the
clusters with the highest extinction estimates in Table \ref{basics.tab},
[H94a]4 and complex 6+7, are located in regions with higher-than-average
patchy extinction.

\section{Starburst time-scales}
\label{context.sec}

\subsection{[Fe {\sc ii}] vs. Brackett line emission}
\label{context1.sec}

In Section \ref{cirpass.sec} we showed that while the [Fe {\sc ii}]
line emission roughly follows the distribution of the YMCs across the
face of NGC 1140, the high-order Brackett lines are predominantly
confined to the northern starburst region. The luminosity of this
region is, at optical continuum wavelengths, dominated by a complex of
three of the most massive compact YMCs in the galaxy, [H94a]1--3.

The [Fe {\sc ii}] line on the one hand, and the Brackett 12--4 and 14--4 lines
on the other originate in physically distinct processes, each governed by
their unique time-scale(s). It is well-established that [Fe {\sc ii}]$\lambda
1.64 \mu$m line emission originates through two different mechanisms. Strong,
compact [Fe {\sc ii}] emission originates in partially ionised zones or
shock-excited gas produced by SNe. In addition, more diffuse, spatially
extended [Fe {\sc ii}] emission seems to be associated with galactic
superwinds. In active galactic nuclei, containing large numbers of partially
ionised clouds and a strong photo-ionising source, [Fe {\sc ii}] emission
originates in the partially ionised zones (see Thompson [1995] and references
therein), while in starburst galaxies, compact [Fe {\sc ii}] emission is
thought to arise from discrete SNRs (e.g., Oliva, Moorwood \& Danziger 1989,
Lumsden \& Puxley 1995, Vanzi \& Rieke 1997, Morel, Doyon \& St-Louis 2002,
Alonso-Herrero et al. 2003). Its excitation requires both the destruction of
dust grains (which contain a large fraction of the interstellar iron) {\it
and} large transition zones between H{\sc ii} regions and regions of neutral
hydrogen, in association with a hard ionising source (so that the electron
temperature is sufficiently high to excite the forbidden [Fe {\sc ii}]
transitions; e.g., Oliva et al. 1989). Such spatial scales therefore imply
that thermal shock excitation is favoured (Oliva et al. 1989, Vanzi \& Rieke
1997).

Hydrogen Brackett lines, on the other hand, are formed by
recombination in ionised gas associated with sources having a strong
Lyman continuum, such as H{\sc ii} regions created by young, massive
stars. The [Fe {\sc ii}]$\lambda 1.64 \mu$m/Br$\gamma$ ratio is
therefore a good diagnostic to distinguish between the ionisation
structures favouring either ``fast'' shock-driven (as in SNRs) or UV
destruction (as in H{\sc ii} regions) excitation (see, e.g., Oliva et
al. 1989, Vanzi \& Rieke 1997 and references therein). The strong
(high-order) Brackett line emission will fade effectively after $\sim
8$ Myr, while the [Fe {\sc ii}] emission remains observable
significantly longer; the [Fe {\sc ii}]/Brackett line ratio is
therefore expected to be age dependent.

This difference in origin is quantitatively supported by our line strength
measurements. While both the composite Br 12--4 and Br 14--4 emission lines in
NGC 1140-N are unresolved (with measured FWHMs of $3.5\pm0.2${\AA} and
$3.2\pm0.2${\AA} for Br 12--4 and Br 14--4 respectively, compared to the
widths of unresolved sky lines of FWHM $3.8\pm0.2$\AA), the [Fe {\sc ii}]
lines are clearly resolved. We measure FWHMs of $(7.0 \pm 0.2)${\AA} and $(5.7
\pm 0.2)${\AA} in starburst regions S and N, respectively, corresponding to
intrinsic linewidths of 5.9{\AA} and 4.2{\AA} (FWHM), after correcting for the
widths of the unresolved lines. This implies that the velocity dispersions
across regions S and N are $\sim 75$ and $\sim 110$ km s$^{-1}$, respectively,
which is consistent with the scenario that we are seeing outflows in our [Fe
{\sc ii}] emission, most likely driven by the combination of SNe and stellar
winds associated with the starburst.

It has been suggested that the [Fe {\sc ii}] line emission reaches a maximum
luminosity when the expansion of the entire SNR (of SN type Ia) becomes
radiative after its initial adiabatic phase (Lumsden \& Puxley 1995, Morel et
al. 2002; see also Shull \& Draine 1987). This occurs some $10^4$ yr after the
SN explosion, when radiative cooling occurs on shorter time-scales than the
dynamical time-scale. The presence of strong [Fe {\sc ii}] emission indicative
of SNRs can therefore help to set limits on a galaxy's star formation history.
The last SNe in a quenched starburst region occur at a time comparable to the
longest lifetime of an SN progenitor after the end of the starburst activity.
Following Iben \& Laughlin (1989) and Hansen \& Kawaler (1994), the time $t$
spent between the zero-age main sequence and planetary nebula phase by an $8
{\rm M}_\odot$ progenitor star, which is generally adopted as a lower limit
for Type II SNe (e.g., Kennicutt 1984), corresponds to $t \sim 35-55$ Myr.

Thus, for NGC 1140 we conclude that the presence of strong [Fe {\sc ii}]
emission throughout the central galaxy implies that most of the recent
star-formation activity was induced some $35-55$ Myr ago. However, based on
our CIRPASS spectroscopy alone, we cannot exclude the possibility that much of
the unresolved [Fe {\sc ii}] emission outside the distinct peaks is associated
with older, dissolved SNRs that have expanded into the galaxy's ISM in the
mean time.

The Br 12--4 and Br 14--4 emission, concentrated around NGC 1140-N, is
most likely associated with recombination processes in (younger,
ionised) H{\sc ii} regions. This implies, therefore, that the most
recent starburst event has transpired in this more confined area of
the NGC 1140 disc. Thus, based on the [Fe {\sc ii}] versus Brackett
line emission, we conclude that a galaxy-wide starburst was induced
several tens of Myr ago, with more recent starburst activity
concentrated around the northern starburst region, NGC 1140-N, and the
massive, compact YMCs [H94a]1--3. This is quantitatively supported by
the stronger CO(6,3) absorption feature detected in this region,
compared to the southern starburst, which again implies that this
emission is between about 7 and 10 Myr old, if it is indeed associated
with the presence of RSG stars. In fact, since the high-order Brackett
lines will essentially disappear after about 8 Myr, the presence of
both the Brackett lines and the CO(6,3) absorption feature places
tight constraints on the age of NGC 1140-N.

The scenario sketched above is quantitatively and qualitatively
confirmed by our results based on broad-band {\sl HST} imaging
observations. There appears to be a small age difference between YMC
complexes [H94a]1--3 and [H94a]6--7 in the sense expected from our
CIRPASS analysis (although we should keep in mind the caveats related
to the broad-band analysis that were discussed in detail in Section
\ref{hst2.sec}).

\begin{table}
\caption[ ]{\label{fluxes.tab}Fluxes and equivalent width measurements
of NGC 1140-N and S}
\begin{center}
\begin{tabular}{lccc}
\hline
\hline
\multicolumn{1}{c}{Feature} & \multicolumn{1}{c}{$F_{\rm tot} (\times
10^{-17})$} & \multicolumn{1}{c}{$F_{\rm peak} (\times 10^{-17})$} &
\multicolumn{1}{c}{EW$^a$} \\
 & \multicolumn{1}{c}{(erg s$^{-1}$ cm$^{-2}$)} & \multicolumn{1}{c}{(erg
 s$^{-1}$ cm$^{-2}$ {\AA}$^{-1}$)} & \multicolumn{1}{c}{({\AA})} \\
\hline 
\multicolumn{3}{c}{1. NGC 1140-N} \\

[Fe {\sc ii}]   & 48 & 1.1 & 4 \\
Br 12--4        & 13 & 1.2 & 2 \\
Br 14--4        & 28 & 1.2 & 1 \\
CO              &    &     & 5 \\
Continuum       &    & 0.9 \\
\\
\multicolumn{3}{c}{2. NGC 1140-S} \\

[Fe {\sc ii}]   & 56 & 0.8 & 8 \\
Continuum       &    & 0.3 & \\
\hline
\end{tabular}
\end{center}
{\sc Note} -- $^a$ The uncertain continuum level, in part caused by
the underlying stellar absorption features
(cf. Fig. \ref{spectra.fig}) -- such as the CO(4,1), CO(5,2), CO(6,3)
and CO(7,4) bands, which will be present in a stellar population as
soon as the supergiant or giant stars appear; OH absorption is usually
less important (cf. Lan\c{c}on \& Wood 2000) -- implies that the
inherent uncertainties of our EW measurements are on the order of
30--50 per cent, in particular for ages $\gtrsim 7$ Myr.
\end{table}

As a final check of the validity of our arguments we refer to the flux
measurements reported in the various sections of this paper, an
overview of which is given in Table \ref{fluxes.tab}. In principle,
our detection of a possible age gradient across the face of NGC 1140
could have been introduced artificially, i.e., by the limited
sensitivity of the IFU, in particular in region NGC 1140-S. The
crucial test is therefore to determine whether we should have been
able to detect {\it any} flux from the high-order Brackett lines in
NGC 1140-S if the ratio of the [Fe {\sc ii}] line flux to that of the
Br 12--4 and Br 14--4 lines were roughly constant from NGC 1140-N to
S. From the previous discussion, the measurements collected in Table
\ref{fluxes.tab}, and from Fig. \ref{emission.fig}, it follows that we
did not detect {\it any} integrated Br 12--4 or 14--4 flux in the
southern starburst region, above a $\sim 3 \sigma$ noise level in the
individual fibre spectra. However, by itself this does not mean that
the ratios between regions N and S must be different per se. In fact,
if we consider the peak fluxes in NGC 1140-N, the ratio of [Fe {\sc
ii}] to either of the Brackett lines is close to unity within the
uncertainties (which are on the order of 30 per cent for all
measurements), but their integrated flux ratios are significantly
higher. This indicates that, while the peak ratios come from the same
set of fibres, the overall distribution of the Brackett line emission
in NGC 1140-N is less extended spatially than that of the [Fe {\sc
ii}] emission.

As already briefly referred to, from Figs. \ref{emission.fig}c and d
it follows that in neither of the Br 12--4 nor the Br 14--4 lines we
detect any flux above our detection level, in any of the fibres in
region NGC 1140-S.  This includes the fibres coincident with the peak
flux measurement in the [Fe {\sc ii}] line. If, however, as in region
N, the peak flux ratios would also be close to unity in NGC 1140-S, we
should have detected at least high-order Brackett line emission at the
peak position, because the [Fe {\sc ii}] emission in the NGC 1140-S
peak is significant in terms of the S/N ratio.  Therefore, this
``sanity check'' implies that the [Fe {\sc ii}] to Brackett line ratio
is a function of position in the NGC 1140 core, and thus it seems
likely that the suggested age gradient is indeed real.

We therefore conclude that there is no region of intense, current star
formation in NGC 1140-S similar to the active starburst area observed in NGC
1140-N. The star formation level must therefore have declined in region S
compared to that in N.

\subsection{Comparison with independent determinations}
\label{context2.sec}

Based on their preliminary analysis of the star cluster population in
NGC 1140, H94a concluded that the accretion of a companion galaxy
triggered an extended starburst and subsquent compact YMC formation in
the core of the merger remnant. Stellar population synthesis of both
an {\it IUE} UV spectrum of the central region (suggesting a
population of $\sim 10^4$ OB stars) and of {\sl HST} optical colours
(and by analogy to [compact] YMC properties in a number of other
starburst galaxies) indicates that the clusters [H94a]1--4 might be
$\sim 3$ Myr old, while clusters [H94a]6 and 7, on the other end of
the central high-intensity region, are somewhat older, at $\sim 15$
Myr. In addition, a blue arclet extending outward from clusters 6 and
7 contains a string of smaller H{\sc ii} regions or 30
Doradus/R136-type star clusters, with slightly redder colours than the
compact YMCs, which are likely older but otherwise similar objects. A
small age gradient appears therefore to be present throughout the
galaxy's central region. Finally, relatively blue optical colours in
the galaxy's outer regions -- where also redder, slightly less
luminous star clusters are found -- suggest that extensive star
formation occurred throughout NGC 1140 within the last $\sim 1$ Gyr
(H94a); current star formation could possibly be sustained for $\sim
6$ Gyr, based on the available gas mass and the global star-formation
rate (H94b). We note that while we are able to confirm H94a's age
estimates of clusters [H94a]6 and 7 to within the uncertainties, we
cannot confirm the very young ages postulated for clusters 1--3 based
on our broad-band SED analysis (but see Fig. \ref{ccdiag.fig}, which
suggests that complex [H94a]1--3 might indeed be younger than 4 Myr)
due to the technical limitation that our minimum model age is 4 Myr
(and the fact that our models include uncertain descriptions of the
poorly-understood RSG stellar evolutionary phase, and no description
of the evolution of Wolf-Rayet stars).  Nevertheless, our {\sl HST}
results are consistent with the age difference implied by the H94a
analysis.

The time-scales derived for both the duration of the starburst event induced
in the centre of NGC 1140, and also for substantial fluctuations of the star
formation rate during the starburst, based on both our CIRPASS spectroscopy,
and independently confirmed by our fits to the broad-band SEDs of the few
dozen compact YMCs, are consistent with the detailed analyses of the stellar
population composition in the disc of NGC 1140 by both Takagi et al. (1999)
and Cid Fernandez, Le\~ao \& Rodrigues Lacerda (2003), based on broad-band
SEDs and high-resolution optical spectroscopy, respectively. The latter
authors derive a mean starburst age of $\langle\log(t_{\rm SB} / {\rm
yr})\rangle = 7.2 \pm 0.6$, or $\langle t_{\rm SB} \rangle \sim 16$ Myr; they
argue that almost half ($46 \pm 6$ per cent) of the disc stellar population is
younger than a few $\times 10^7$ yr, with a further $35 \pm 7$ per cent being
as young as a few $\times 10^8$ yr.

Takagi et al.'s (1999) estimates concur with these results. Their starburst
age estimates range from 25 to 40 Myr, depending on both the extinction law
assumed and the mass fraction of the galaxy involved in the starburst. In
their most recent analysis (Takagi et al. 2003), however, they derive a
starburst duration of about 300 Myr, based on an analysis of UV--optical SEDs
and allowing for additional gas infall over the galaxy's lifetime. Their age
derivation is, however, significantly affected by the age--extinction
degeneracy, which is of particular importance around the starburst age adopted
for NGC 1140.

\section{NGC 1140's current supernova rate}
\label{snr.sec}

Significant effort has been invested in the use of the [Fe {\sc ii}]
luminosity of a given galaxy as indicator of its overall SN rate (e.g.,
Calzetti 1997, Vanzi \& Rieke 1997, Alonso-Herrero et al. 2003 and references
therein). Although our CIRPASS observations were obtained under
non-photometric conditions, we can still derive a rough SN rate based on the
[Fe {\sc ii}] line flux. To do so, we used standard star observations taken on
a different night, which implies that the calibration does not take into
account the differences in transparency of the atmosphere between both nights.
Since our observations were taken under non-photometric conditions, this
essentially implies that we can at best obtain a lower limit to the galaxy's
SN rate. 

The median combined [Fe {\sc ii}] flux of the southern starburst region NGC
1140-S is $F_{\rm [Fe {\sc ii}],S} \gtrsim 56 \times 10^{-17}$ ergs cm$^{-2}$
s$^{-1}$. Assuming a distance to NGC 1140 of 20 Mpc, this corresponds to a
luminosity of $L_{\rm [Fe {\sc ii}],S} \gtrsim 27 \times 10^{36}$ ergs
s$^{-1}$. For the northern starburst region, the combined [Fe {\sc ii}] flux
of $\gtrsim 48 \times 10^{-17}$ ergs cm$^{-2}$ s$^{-1}$ corresponds to an
integrated luminosity of $L_{\rm [Fe {\sc ii}],N} \gtrsim 23 \times 10^{36}$
ergs s$^{-1}$.

Using Vanzi \& Rieke's (1997) calibration of the relationship between
[Fe {\sc ii}] luminosity and SN rate, based on the SN rate of 0.11 SN
yr$^{-1}$ for the prototype starburst galaxy M82 (Huang et al. 1994),
we derive SN rates of $\gtrsim 0.3$ SN yr$^{-1}$ for both regions. In
line with the $\sim (1.5 - 2) \times$ higher [Fe {\sc ii}] luminosity
for M82 measured by Alonso-Herrero et al. (2003), we arrive at a
roughly double SN rate if we use their calibration.  We caution that
the uncertainties in these SN rates are at least a factor of two,
since such are the uncertainties in M82's SN rate itself. It thus
follows that our derived SN rates are of a similar order of magnitude
as those in the active starburst nucleus of M82. However, since our
CO(6,3) absorption feature is weaker than that seen in M82
(cf. F\"orster-Schreiber et al. 2001, their Fig. 2), the NGC 1140
starburst is likely somewhat older than the one in M82's active
centre.

For comparison, Calzetti (1997) predicted SN rates -- based on the long-slit
NIR spectra of Calzetti, Kinney \& Storchi-Bergmann (1996) -- for the very
central region of NGC 1140-N ranging from $0.008 - 0.034$ SN yr$^{-1}$ (i.e.,
roughly an order of magnitude lower than ours) for stars with minimum mass of
$8 {\rm M}_\odot$, based on the current star-formation rates obtained from
Br$\gamma$ luminosities and for upper masses to the IMF of 100 and $30 {\rm
M}_\odot$, respectively. The significant difference in the SN rates derived
from Calzetti's (1997) spectra (which cover only the very central section of
the NGC 1140-N starburst region), compared to the SN rate we derive from our
CIRPASS data covering a larger area imply, therefore, that the SN activity
associated with the active starburst in NGC 1140-N is significantly extended
on spatial scales and not just confined to the area of the highest-intensity
[Fe {\sc ii}] emission.

From a comparison of the integrated [Fe {\sc ii}] luminosities in regions S
and N, it follows that the strength of the starburst is similar in both
regions. The integrated flux of the N+S starburst activity, as traced by the
[Fe {\sc ii}] emission, $F_{\rm [Fe {\sc ii}],N+S} \gtrsim 112 \times
10^{-17}$ ergs cm$^{-2}$ s$^{-1}$, comprises $\simeq 40$ per cent of the total
[Fe {\sc ii}] flux within our common CIRPASS/{\sl HST} FoV\footnote{Note that
the sum of the [Fe {\sc ii}] fluxes determined in regions S and N is $\sim 9$
per cent smaller than the integrated flux in regions S and N taken together,
and determined independently. This difference is most likely due to
uncertainties in the [Fe {\sc ii}] line profile fits, in particular at low S/N
ratios. This also implies that, in addition to the uncertainties introduced by
the relative photometry, our {\it systematic} uncertainty in the [Fe {\sc ii}]
fluxes and luminosities and the derived SN rates is on the order of (at least)
10 per cent.}: the integrated flux from the 177 CIRPASS lenses covering the
{\sl HST} FoV corresponding to that shown in Fig. \ref{emission.fig}a is
$F_{\rm [Fe {\sc ii}],tot} \gtrsim 285 \times 10^{-17}$ ergs cm$^{-2}$
s$^{-1}$. Thus, our CIRPASS observations have revealed the presence of a large
fraction ($\sim 60$ per cent) of diffuse [Fe {\sc ii}] emission throughout the
entire centre of NGC 1140. For the high-order Brackett lines the situation
is very different indeed; there is a negligible fraction of Br 12--4 and Br
14--4 flux present outside the NGC 1140-N starburst site, which further
supports our interpretation that these lines are associated with the most
active star formation in the galaxy.

\section{Summary and conclusions}
\label{summary.sec}

We have analysed NIR integral field spectroscopy of the nearby
southern starburst galaxy NGC 1140, obtained at the Gemini-South
telescope equipped with CIRPASS. The wavelength coverage used for this
programme, $0.22 \mu$m from $\sim 1.45 - 1.67 \mu$m, includes the
bright [Fe {\sc ii}]$\lambda 1.64 \mu$m emission line, as well as the
high-order Brackett lines Br 12--4 and Br 14--4, which are all clearly
resolved in the NGC 1140 central starburst region, and located well
away from contaminating OH sky lines. Additional {\sl HST} imaging
observations covering the UV/optical range have been used to place
these CIRPASS results in the context of interaction-induced star
cluster formation and evolution.

NGC 1140 is remarkable in the number of compact YMCs formed in its bright,
compact emission-line core, which were most likely triggered by a recent
merger event with a gas-rich companion galaxy. The total energy output of
these YMCs far exceeds that of the local ``super'' star cluster R136 in the
core of the 30 Doradus starburst region in the LMC. Our CIRPASS spectroscopy,
combined with the broad-band {\sl HST} images, probes the complex transitions
between the YMCs and the galaxy's ISM, which we use to test predictions of
models for radiative and supernova-induced shocks.

The flux distribution of the [Fe {\sc ii}] line is clearly more extended than
those of the Brackett lines. While [Fe {\sc ii}] emission is found throughout
the galaxy, both Br 12--4 and Br 14--4 emission is predominantly associated
with the northern starburst region. It is now well-established that the {\it
strong} [Fe {\sc ii}]$\lambda 1.64 \mu$m line emission in starbursts
originates predominantly in SNRs. The Brackett lines, on the other hand, are
formed by recombination in ionised gas associated with sources having a strong
Lyman continuum, such as H{\sc ii} regions created by young, massive stars.
The [Fe {\sc ii}] emission and the young region as a whole may, in fact,
represent the broad base of a galactic outflow, or wind, in its early stages.

The presence of strong [Fe {\sc ii}] emission in NGC 1140's core, indicative
of SNRs, can help to set limits on its star formation history. The last SNe in
a quenched starburst region occur at a time comparable to the longest lifetime
of an SN progenitor after the end of the starburst activity. Thus, for NGC
1140 we conclude that the presence of strong [Fe {\sc ii}] emission throughout
the central galaxy implies that most of the recent star-formation activity was
induced in the past $\sim 35-55$ Myr.

The Br 12--4 and Br 14--4 emission, concentrated around NGC 1140-N, is most
likely associated with recombination in (younger) H{\sc ii} regions. This
implies, therefore, that the most recent starburst event has transpired in
this more confined area of the NGC 1140 disc, around the northern starburst
region and the massive, compact YMCs [H94a]1--3.

This scenario is provisionally confirmed by our fits based on
broad-band {\sl HST} imaging observations (despite the large
associated uncertainties), and by the preliminary analysis of
H94a. Using the method developed in Anders et al. (2004a) and de Grijs
et al. (2003c), we derive the YMC ages, masses, and extinction values
simultaneously from the {\sl HST} broad-band passband coverage; we
assume a generic metallicity for all NGC 1140 YMCs of $0.4 Z_\odot$,
which we justify by emphasizing technical limitations of our
metallicity fits at the young ages expected for the compact NGC 1140
YMCs. Such a metallicity is also supported by previous, independent
estimates. The ages derived for the NGC 1140 YMCs are all $\lesssim
20$ Myr (or perhaps significantly younger, in view of the
uncertainties inherent to our method of analysis of broad-band SEDs),
consistent with independently determined estimates of the galaxy's
starburst age, while there appears to be an age difference between YMC
complexes [H94a]1--3 and [H94a]6--7 in the sense expected from our
CIRPASS analysis. Our photometric mass estimates of the NGC 1140 YMCs,
likely representing upper limits, are comparable to the Galactic GCs
comprising the high-mass wing of the GC mass distribution and to the
spectroscopically confirmed masses of the compact YMCs in M82 and the
Antennae galaxies. Our detection of similarly massive YMCs in NGC 1140
supports the scenario that such objects form preferentially in the
extreme environments of interacting and starburst galaxies.

Finally, we derive lower limits to the SN rates in the NGC 1140 starburst
regions S and N of $\sim 0.3$ SN yr$^{-1}$, within a factor of $\sim 2$, which
implies a similar starburst vigour as seen in the actively starbursting
nucleus of the nearby prototype starburst galaxy M82. The two
spatially-confined starburst sites S and N emit $\sim 40$ per cent of the
integrated [Fe {\sc ii}] flux in the galaxy, with the remainder being
distributed in a diffuse component. The high-order Brackett lines, on the
other hand, exhibit a negligible diffuse component, but are confined to
starburst region N.

\section*{Acknowledgments} This paper is partially based on observations
obtained at the Gemini Observatory, which is operated by the Association of
Universities for Research in Astronomy, Inc. (AURA), under a cooperative
agreement with the U.S. National Science Foundation (NSF) on behalf of the
Gemini partnership: the Particle Physics and Astronomy Research Council
(PPARC, UK), the NSF (USA), the National Research Council (Canada), CONICYT
(Chile), the Australian Research Council (Australia), CNPq (Brazil) and
CONICET (Argentina). We are grateful to Matt Mountain for the Director's
discretionary time to demonstrate the scientific potential of integral field
units (the PI's of this demonstration science programme are Andrew Bunker,
Gerry Gilmore, and Roger Davies). We thank the Gemini Board and the Gemini
Science Committee for the opportunity to commission CIRPASS on the
Gemini-South telescope as a visitor instrument. We thank Phil Puxley, Jean
Ren\'e-Roy, Doug Simons, Bryan Miller, Tom Hayward, Bernadette Rodgers, Gelys
Trancho, Marie-Claire Hainaut-Rouelle and James Turner for the excellent
support received. CIRPASS was built by the instrumentation group of the
Institute of Astronomy in Cambridge, UK. We warmly thank the Raymond and
Beverly Sackler Foundation and PPARC for funding this project. Andrew Dean,
Anamparambu Ramaprakash and Anthony Horton all assisted with the observations
in Chile, and we are indebted to Dave King, Jim Pritchard \& Steve Medlen for
contributing their instrument expertise. The optimal extraction software for
this 3D fibre spectroscopy was written by Rachel Johnson and Andrew Dean. This
research is also partially based on observations with the NASA/ESA {\sl Hubble
Space Telescope}, obtained at the Space Telescope Science Institute (STScI),
which is operated by AURA under NASA contract NAS 5-26555. JSG thanks the
University of Wisconsin Graduate School for research support, and the
Department of Physics and Astronomy at University College London for their
hospitality for work on this paper. This research has made use of NASA's
Astrophysics Data System Abstract Service.

\end{document}